\documentclass[12pt]{article} 
\usepackage[utf8]{inputenc} 
 
\usepackage{hyperref}
\hypersetup{hypertexnames=false, colorlinks=true, citecolor=blue,urlcolor=blue,filecolor=red}
 
 \usepackage{geometry} 
 \usepackage{graphicx} 
 
 \usepackage{booktabs} 
 \usepackage{array} 
 \usepackage{paralist} 
 \usepackage{verbatim} 
 \usepackage{subfig} 
 \usepackage{amsbsy}
 \usepackage{amsmath}
 \usepackage{amssymb}
 \usepackage{float}
 \usepackage[toc,page,title]{appendix}
 \usepackage{lineno}
 \usepackage{mathrsfs}
 \usepackage{lineno}
 \usepackage{todonotes}
 \usepackage{setspace}
 \usepackage{natbib}
 \usepackage{multirow}
 \usepackage{textcomp}
 \usepackage{data_values}
 \usepackage{missing_traits}
 \usepackage{makecell}
 \usepackage{refcount}

 \usepackage{fancyhdr} 
 \usepackage{titlesec}
 \titleformat{\section}{\sffamily\mdseries\upshape\uppercase}{\thesection.}{1em}{}
 \titleformat{\subsection}{\sffamily\mdseries\upshape}{\thesubsection}{1em}{}
 \titleformat{\subsubsection}{\sffamily\mdseries\upshape}{\thesubsubsection}{1em}{}

 \usepackage[nottoc,notlof,notlot]{tocbibind} 
 \usepackage[titles,subfigure]{tocloft} 
 
 \setcitestyle{aysep={}}
 
 \graphicspath{{./}}
 
 \newif\ifblinded
 \newcommand{\blind}[1]{\ifblinded {[blinded]} \else #1 \fi}
 
 \blindedfalse 

\begin{document}
		\pagenumbering{gobble}
	\begin{flushright}
		Version dated: \today
	\end{flushright}
	
	\begin{center}
		
		\centerline{\Large \bf Inferring Phenotypic Trait Evolution on Large Trees}
		\centerline{\Large \bf With Many Incomplete Measurements}
		\bigskip
		\doublespacing
		
		\blind{
			\noindent{\normalsize \sc
				Gabriel Hassler$^1$, \\
				Max R.~Tolkoff$^2$, \\
				William L. Allen$^{3}$, \\
				Lam Si Tung Ho$^{4}$, \\
				Philippe Lemey$^{5}$, \\
				and Marc A.~Suchard$^{1,2,6}$}\\}
		
		\singlespacing
		\bigskip
		\blind{
			\noindent {\small
				\it $^1$Department of Biomathematics, David Geffen School of Medicine at UCLA, University of California,
				Los Angeles, United States \\
				\medskip
				\it $^2$Department of Biostatistics, Jonathan and Karin Fielding School of Public Health, University
				of California, Los Angeles, United States} \\
				\medskip
			\it $^3$Department of Biosciences, Swansea University, Swansea, United Kingdom \\
			\medskip
			\it $^4$Department of Mathematics and Statistics, Dalhousie University, Halifax, Nova Scotia, Canada \\
			\medskip
			\it $^5$Department of Microbiology and Immunology, Rega Institute, KU Leuven, Leuven, Belgium \\
			\medskip
			\it $^6$Department of Human Genetics, David Geffen School of Medicine at UCLA, Universtiy of California,
			Los Angeles, United States}
		
	\end{center}

	\noindent{\bf Corresponding author:} Marc A. Suchard, Departments of Biostatistics, Biomathematics, and Human Genetics,
	University of California, Los Angeles, 695 Charles E. Young Dr., South,
	Los Angeles, CA 90095-7088, USA; E-mail: \url{msuchard@ucla.edu}\\
	\clearpage

	\pagenumbering{arabic}

	%
	
	\clearpage
	
		\onehalfspacing
	
	\paragraph{Abstract}
	Comparative biologists are often interested in inferring covariation between multiple biological traits sampled across numerous related taxa.
	To properly study these relationships, we must control for the shared evolutionary history of the taxa to avoid spurious inference.
	Existing control techniques almost universally scale poorly as the number of taxa increases.
	An additional challenge arises as obtaining a full suite of measurements becomes increasingly difficult with increasing taxa.
	This typically necessitates data imputation or integration that further exacerbates scalability.
	We propose an inference technique that integrates out missing measurements analytically and scales linearly with the number of taxa by using a post-order traversal algorithm under a multivariate Brownian diffusion (MBD) model to characterize	 trait evolution.
	We further exploit this technique to extend the MBD model to account for sampling error or non-heritable residual variance.
	We test these methods to examine mammalian life history traits, prokaryotic genomic and phenotypic traits, and HIV infection traits.
	We find computational efficiency increases that top two orders-of-magnitude over current best practices.
	While we focus on the utility of this algorithm in phylogenetic comparative methods, our approach generalizes to solve long-standing challenges in computing the likelihood for matrix-normal and multivariate normal distributions with missing data at scale.

	
	\paragraph{Keywords:} Bayesian inference, matrix-normal, missing data, phylogenetics

	\label{ch:PFA}
	\section{Introduction}
	
	Phylogenetic comparative methods explore the relationships between different biological phenotypes across sets of organisms.
	To properly understand these phenotypic trait relationships, methods must adjust for the shared evolutionary history of the taxa \citep{Felsenstein85}.
	Molecular sequences from emerging sequencing technology and high-throughput biological experimentation enable such phylogenetic adjustment for rapidly growing numbers of taxa and increasing numbers of trait measurements.
	Comparative studies incorporating dense taxonomic sampling create the potential for new research into general patterns in phenotypic evolution, key differences between subgroups and the relationship between phenotypic and genetic evolutionary dynamics.
	Unfortunately, many phylogenetic comparative methods remain poorly equipped to handle these research questions at scale.
	
	Popular methods often assume an underlying Brownian diffusion process acts along each branch of a phylogenetic tree, such that the traits are multivariate normally distributed.
	\cite{revell2012} and \cite{Adams14PLGS}, for example, parameterize this distribution in terms of a highly-structured variance-covariance matrix that characterizes the tree and trait covariation.
	Computational work to invert this matrix to evaluate the multivariate normal likelihood scales cubically with the number of taxa.
	This work stands even more troublesome when the phylogenetic tree remains unknown and requires joint inference with the trait process, necessitating repeated inversion.
	\cite{freckleton2012fast}, \cite{Pybus2012}, and \cite{ho2014linear} all independently develop algorithms that take advantage of the matrix-normal structure of the data under the MBD model to evaluate the likelihood.
	Using the tree structure, these algorithms then scale linearly with the number of taxa with complete data, but this ideal run-time currently stumbles when trait measurements are missing.
	
	As the number of taxa grows large, measuring a complete suite of traits for all taxa becomes increasingly challenging.
	Recent solutions to this problem in phylogenetics include those proposed by \cite{MaxPaper1}, who use Markov chain Monte Carlo (MCMC) to numerically integrate via sampling the missing data, and \cite{bastide2018inference}, who exploit expectation-maximization (EM) to impute missing values and all unobserved ancestral node traits.
	Both lines of work, however, require iterative manipulation of the likelihood function on a per-taxon basis.
	Current sampling methods scale at best quadratically with the number of taxa, and likewise the number of ancestral node trait values that require imputation is directly related to the number of taxa.
	As a consequence, these methods remain computationally prohibitive for large trees.
	
	In this paper, we reformulate evaluation of the data likelihood function under a Brownian diffusion process on a tree such that we achieve the marginalized likelihood of the observed trait measurements only.
	This innovation arises from thinking about observed tip traits as multivariate normally distributed with infinite precision in their sampling, while missing traits have zero precision, and appropriately propagating these precisions up the tree through dynamic programming involving an unusual matrix pseudo-inverse definition.
	This pseudo-inverse finds similar use, but independent discovery, in \cite{bastide2018inference}.
	Unlike previous approaches, the integration avoids EM iteration making simultaneous inference with the phylogeny practical and enables researchers to analyze all available measurements when inferring the trait relationships.
	Surprisingly, we can still evaluate the observed-data likelihood in linear time with respect to the number of taxa.
	The price to be paid is that computation now scales cubically, rather than quadratically, in the number of traits.
	This remains a small price since the number of taxa is often orders-of-magnitude larger than the number of traits.
	It is also notable that this method has applications beyond phylogenetic comparative methods and can be used more generally in a special class of matrix-normal and multivariate normal distributions with missing data.
	This has been a long standing problem in statistics since at least the 1930's \citep{wilks1932moments}, with more recent work by \cite{dominici2000conjugate}; \cite{cantet2004full}; \cite{allen2010transposable}; and \cite{glanz2013expectation}. One important limitation to our approach is that it assumes data are missing at random \citep{little1987statistical} which is inappropriate for many data sets.
	
	We also demonstrate how this framework can be easily extended 
	to incorporate residual variance in the MBD model, which is only one of many possible model extensions.
	Our strategy of analytically marginalizing the observed data likelihood extends seamlessly to this and other model extensions and allows for efficient inference on these models while maintaining likelihood computations that scale linearly with the number of taxa.
	These extensions open up lines of inquiry not available in the simple MBD model.
	In particular, including residual variance in the model enables inference of phylogenetic heritability.
	
	We demonstrate the broad utility of our algorithm to compute the marginalized likelihood through three examples.
	First, we examine covariation in mammalian life history traits using data on 3690 taxa from the PanTHERIA ecological database \citep{Jones2009}.
	Second, we use our new efficient algorithm to simultaneously evaluate several theories regarding prokaryotic evolutionary theory.
	We use data from NCBI Genome and a recent study by \cite{goberna2016predicting}, along with matching 16S sequences from the ARB Silva Database \citep{ludwig2004arb}, to jointly infer both the phylogenetic tree and evolutionary correlation between several prokaryotic genotypic and phenotypic traits.
	Finally, we apply our multivariate residual variance model extension to data presented by \cite{blanquart17Heritability} concerning HIV virulence to evaluate the heritability of HIV viral load and CD4 T-cell decline.
	We compare the computation speed of our analytical integration method against current best-practice methods and observed increases in speed that top two orders-of-magnitude.

	%

	
	%

	%
	
	\section{Phenotypic diffusion on trees} \label{sec:diffusion}

	Consider a data-complete collection $\datamat = \left(\datamat_{1}, \ldots,\datamat_{\ntaxa} \right)\transpose$ where $\datamat_{\taxonIdx} = \left(\datamatsm_{\taxonIdx 1}, \ldots, \datamatsm_{\taxonIdx \ntraits} \right)\transpose$ of $\ntraits$ real-valued phenotypic traits measured across $\ntaxa$ biological taxa.
	Relating the taxa stands a known and fixed or unknown and random phylogeny $\treeparam$ that is a bifurcating, directed acyclic graph whose $2\ntaxa - 1$ vertices originate with a degree-2 root node $\node_{2 \ntaxa - 1}$ and terminate with degree-1 tip nodes $\left(\node_1, \ldots, \node_{\ntaxa} \right)$ that correspond to the $\ntaxa$ taxa.
	Linking vertices are edge weights or branch lengths $(t_1, \ldots, t_{2 \ntaxa - 2})$. Let $\trdatamat_\nodeIndexThree = \left(\trdatamatsm_{k1}, \hdots, \trdatamatsm_{k\ntraits}\right)$ be latent values of the traits at node $\node_\nodeIndexThree$ on the tree for $\nodeIndexThree = 1, \ldots, 2\ntaxa - 1$.
	For tip nodes $\taxonIdx = 1, \hdots, \ntaxa$, we posit a stochastic link $\stochLink$ where $\datamat_{\taxonIdx}$ is drawn from some distribution parameterized by $\trdatamat_{\taxonIdx}$ and other hyperparameters (see Figure \ref{fig:tikz}).
	Comparative methods standardly assume that the density $\stochLink$ is degenerate at $\trdatamat_{\taxonIdx}$ (i.e.~$\datamat_{\taxonIdx} = \trdatamat_{\taxonIdx}$ with probability 1), but we relax this assumption in future sections.
	
	\newcommand{\latentData}{\trdatamat}
	\newcommand{\data}{\datamat}
	\begin{figure}
		\begin{center}
			\scalebox{1.0}{%
				\begin{tikzpicture}
				\shadedraw[ball color=blue!50, opacity=0.2] (0, 0)  circle (0.35) node[opacity=1.0,minimum size=0.6cm,text width=2cm,align=center] (AA)  {$\latentData_{1}$ };
				%
				\shadedraw[ball color=blue!50, opacity=0.2] (2, 0) circle (0.35) node[opacity=1.0,minimum size=0.6cm,text width=2cm,align=center] (BB) {$\latentData_{2}$ };
				%
				\shadedraw[ball color=blue!50, opacity=0.2] (4, 0) circle (0.35) node[opacity=1.0,minimum size=0.6cm,text width=2cm,align=center] (CC) {$\latentData_{3}$ };
				%
				\shadedraw[ball color=green!50, opacity=0.2] (1, 1.5) circle (0.35) node[opacity=1.0,minimum size=0.6cm,text width=2cm,align=center] (DD) {$\latentData_{4}$ };
				%
				\shadedraw[ball color=green!50, opacity=0.2] (2, 3) circle (0.35) node[opacity=1.0,minimum size=0.6cm,text width=2cm,align=center] (EE) {$\latentData_{5}$};
				%
				%
				\draw (AA) -- (DD) node[midway,above left] {$t_1$};
				\draw (BB) -- (DD) node[midway,above right] {$t_2$};
				\draw (DD) -- (EE) node[midway,above left] {$t_4$};
				\draw (CC) -- (EE) node[midway,above right] {$t_5$};
				%
				%
				%
				%
				\shadedraw[ball color=red!50, opacity=0.2]
				(0, -1.5) circle (0.35) node[opacity=1.0,text width=2cm,align=center] (ZZ) {$\data_{1}$ };
				%
				\shadedraw[ball color=red!50, opacity=0.2]
				(2, -1.5) circle (0.35) node[opacity=1.0,text width=2cm,align=center] {$\data_{2}$ };
				%
				\shadedraw[ball color=red!50, opacity=0.2]
				(4.0, -1.5) circle (0.35) node[opacity=1.0,text width=2cm,align=center] {$\data_{3}$ };
				\draw [dashed] (4.0, -0.5) -- (4.0, -1.0);
				
				\draw [dashed] (2, -0.5) -- (2, -1.0) node[midway,xshift=9mm] {\rotatebox{0}{\footnotesize $\stochLink$}};
				
				\draw [dashed] (0.0, -0.5) -- (0.0, -1.0);
				\end{tikzpicture}
			} 
		\end{center}
		\caption{Schematic of diffusion model with stochastic link function. The data $\datamat = \left(\datamat_1, \datamat_2, \datamat_3\right)\transpose$ arise from latent values $\trdatamat_\taxonIdx$ at the tips of the tree via the stochastic link function $\stochLink$ for $\taxonIdx = 1,\hdots, \ntaxa$.}
		\label{fig:tikz}
	\end{figure}
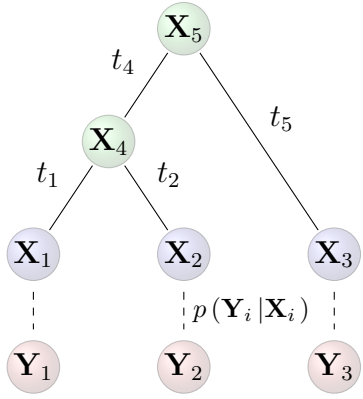
	
	
	The most common phenotypic model of evolution \citep{Felsenstein85} assumes a multivariate Brownian diffusion process acts conditionally independently along each branch generating a multivariate normal (MVN) increment,
	\begin{equation} \label{eq:diffusion_child}
	\trdatamat_{\nodeIndexThree} \sim \mvnDensity{\trdatamat_{\parent{\nodeIndexThree}}}{ t_{\nodeIndexThree} \variance } \text{ for } \nodeIndexThree = 1, \ldots, 2 \ntaxa - 2 ,
	\end{equation}
	centered around the realized value $\trdatamat_{\parent{\nodeIndexThree}}$ at its parent node and variance proportional to an estimable $\ntraits \times \ntraits$ positive-definite matrix $\variance$.
	Since the trait values at the root are also unknown, \cite{Pybus2012} suggest further assuming $\trdatamat_{2 \ntaxa - 1} \sim \mvnDensity{\diffmean}{\pss^{-1} \variance}$ with fixed prior mean $\diffmean$ and sample-size $\pss$.
	

	\subsection{Computation of Observed Data Likelihood} \label{sec:diffusion_likelihood}

	When there are no missing data and under our standard assumption that $\stochLink$ is degenerate, integrating out unobserved internal and root node traits leads to a seemingly simple expression for the data likelihood $\dataLikelihood$ \citep{freckleton2012fast,Vrancken2015}.
	Namely, $\datamat$ is matrix-normal (MN) distributed around mean $\mnMean$, with across-row variance $\treeAndRootVariance$ and across-column variance $\variance$, where $\oneVector{\ntaxa}$ is a vector of length $\ntaxa$ populated by ones, $\oneMatrix{\ntaxa} = \oneVector{\ntaxa}\oneVector{\ntaxa}\transpose$, and $\treeVariance$ is a deterministic function of $\treeparam$.
	Specifically, element $\treeVarianceElement_{\taxonIdx \taxonIdxPrime}$ measures shared evolutionary history and equals the sum of the branch lengths from the root to the most recent common ancestral node of taxa $\taxonIdx$ and $\taxonIdxPrime$ when $\taxonIdx \neq \taxonIdxPrime$ or the sum of the branch lengths from the root to taxon $\taxonIdx$ otherwise. For example, in Figure \ref{fig:tikz}, $\treeVarianceElement_{12} = \branchLength{4}$ and $\treeVarianceElement_{11} = \branchLength{1} + \branchLength{4}$.
	One can evaluate this highly structured matrix-normal likelihood function with computational complexity $\bigo{\ntaxa\ntraits^2}$ given the acyclic nature of $\treeparam$.
	When some data points are missing, however, the observed-data likelihood is no longer matrix-normal and new approaches are needed.
	This becomes increasingly urgent as the prevalence of missing observations grows with the size of trait data sets.
	In this context we wish to compute
	\begin{equation}
	\obsdataLikelihood = \int\dataLikelihoodTwo \differential \datamatmis,
	\end{equation}
	where $\datamatobs$ and $\datamatmis$ contain the observed and missing trait values, respectively.
	
	The two simplest strategies for calculating the observed-data likelihood are, unfortunately, computationally prohibitive for most large problems.
	One such solution forfeits the MN structure of the data in favor a simple expression of the observed-data likelihood.
	This strategy uses the fact that the matrix-normal distribution of $\datamat$ can also be expressed as
	\begin{equation}
	\vecOperator{\dataDist} \sim \mvnDensity{\vecOperator{\mnMean}}{\variance \kronecker \treeVariance},
	\end{equation}	
	using the Kronecker product $\kronecker$.
	Assuming data are missing at random \citep{little1987statistical}, one can simply remove the rows and columns of $\vecOperator{\mnMean}$ and $\variance \kronecker \treeVariance$ corresponding to the missing data and compute the likelihood for this $\ntaxa\ntraits - \nmissingTwo$ dimensional MVN distribution, where $\nmissingTwo$ is the number of missing measurements.
	This likelihood calculation carries the onerous computational complexity $\bigo{\left(\ntaxa\ntraits - \nmissingTwo\right)^3}$.
	Alternatively, from a Bayesian perspective, one could numerically integrate out the missing data by treating each missing data point as an unknown model parameter and employing MCMC to sample each value.
	This strategy restores the matrix-normal structure, but requires the likelihood be evaluated each time one samples a missing data point.
	This results in computation complexity of at least $\bigo{\ntaxa\ntraits^2\nmissing}$, where $\nmissing$ is the number of taxa with missing measurements.
	Because $\nmissing$ often scales with $\ntaxa$, this method remains prohibitively slow for many data sets with large $\ntaxa$.
	Our goal is to integrate out these missing values analytically using a dynamic programming algorithm in order to bring run time down to a much more manageable $\bigo{\ntaxa \ntraits^3}$.

	\subsubsection{Missing Data Definitions and Operations} \label{sec:diffusion_missingops}
	
	To develop our algorithm, we first introduce some useful abstractions and notation.
	At each tip in $\treeparam$, information about each of the $\ntraits$ traits comes in one of three forms: a trait value may be directly observed, latent, or completely missing.
	When directly observed, we posit without loss of generality that the value arises from a normal distribution centered at the observed value  with infinite precision.
	We assume that trait data that arise from latent values are jointly multivariate normally distributed about the unknown latent values with known or estimable precision.
	Finally, a completely missing value arises also without loss of generality from a normal distribution centered at 0 with zero precision.
	To formalize this, for tip $\nodeIndexOne = 1,\ldots,\ntaxa$, we construct a permutation matrix $\nodePermutationMatrix{\nodeIndexOne}$ that groups traits in directly observed, latent, and completely missing order and populate a pseudo-precision matrix
	\begin{equation}
	\nodePrecision{\nodeIndexOne} = \nodePermutationMatrix{\nodeIndexOne}\, \diag{
		\raisedInfty \identityMatrix{}, \tipLatentPrecision{\nodeIndexOne}, 0 \identityMatrix{}
	}
	\nodePermutationMatrix{\nodeIndexOne}\transpose,
	\end{equation}
	where $\diag{\cdot}$ is a function that arranges its constituent elements into block-diagonal form and $\tipLatentPrecision{\nodeIndexOne}$ is the latent block precision.
	Note that any block may be 0-dimensional.
	This construction arbitrarily forces off-diagonal elements of $\nodePrecision{\nodeIndexOne}$ involving directly observed and completely missing traits to equal 0 and plays an important role in simplifying computations. 
	
	We additionally define a series of operations that we will find useful for defining this algorithm.
	We define the pseudo-inverse
	\begin{equation}
	\nodePrecision{\nodeIndexOne}^{\specialInverse} = \nodePermutationMatrix{\nodeIndexOne} \, \diag{
		0 \identityMatrix{}, \tipLatentPrecision{\nodeIndexOne}^{-1}, \raisedInfty \identityMatrix{}
	}\nodePermutationMatrix{\nodeIndexOne}\transpose.
	\end{equation}
	We define the pseudo-determinant $\altDet{}$ as the product of the non-zero singular values.
	We also define the matrix $\notMissingMatrix{\taxonIdx} = \diag{\notMissingIndicator{\taxonIdx 1} , \hdots, \notMissingIndicator{\taxonIdx \ntraits}}$ for $\taxonIdx = 1, \hdots, \ntaxa$, where $\notMissingIndicator{\taxonIdx\traitIdx}$ is an indicator variable which takes a value of $1$ if trait $\datamatsm_{\taxonIdx\traitIdx}$ is observed or latent and $0$ if it is missing.
	Lastly, we define the possibly degenerate multivariate normal density function
	\begin{equation*}
	\log\degmvn{\mvnArg}{\mvnMean}{\mvnPrec}= \frac{1}{2} \log{\altDet{\mvnPrec}}  - \frac{{\effectiveDim{\mvnPrec}}}{2} \log{2 \pi} - {\frac{1}{2} \left(\mvnArg - \mvnMean \right) \transpose \mvnPrec \left( \mvnArg - \mvnMean \right)},
	\end{equation*}
	for some argument $\mvnArg$, mean $\mvnMean$ and precision $\mvnPrec$ of appropriate dimensions.

	\subsubsection{Post-Order Observed Data Likelihood Algorithm} \label{sec:diffusion_postorder}
	Our goal is to efficiently compute the likelihood $\obsdataLikelihood$.
	Following from \cite{Pybus2012}, we perform a post-order traversal where we calculate the observed-data partial likelihood $\cdensity{\datamatobs_{\databelow{\nodeIndexThree}}}{\trdatamat_\nodeIndexThree, \facprecDiff, \treeparam}$ at each node $\node_\nodeIndexThree$ where $\datamatobs_{\databelow{\nodeIndexThree}}$ is the observed data restricted to all descendants of node $\nodeIndexThree$ on the tree.
	For example, in Figure \ref{fig:tikz}, $\datamatobs_{\databelow{4}} = \{\datamatobs_{1}, \datamatobs_{2}\}$.
	
	%
	%
	%
	
	We posit that, given an appropriate stochastic link function $\stochLink$, we can express the observed-data partial likelihood as
	\begin{equation}
	\label{eq:propMVN} \cdensity{\datamatobs_{\databelow{\nodeIndexThree}}}{\trdatamat_\nodeIndexThree, \facprecDiff, \treeparam} =  \remainder{\nodeIndexThree} \degmvn{\trdatamat_{\nodeIndexThree}}{\nodeMean{\nodeIndexThree}}{\nodePrecision{\nodeIndexThree}},
	\end{equation}
	for all nodes $k = 1, \ldots, 2\ntaxa - 1$ and some remainder $\remainder{\nodeIndexThree}$, mean $\nodeMean{\nodeIndexThree}$, and precision $\nodePrecision{\nodeIndexThree}$. Given a parent node $\parentIdx$ with children $\childIdxOne$ and $\childIdxTwo$, let us assume we can express the observed-data likelihood of $\datamatobs_{\databelow{\childIdxOne}}$ and $\datamatobs_{\databelow{\childIdxTwo}}$ as in Equation \ref{eq:propMVN}. Conditioning on $\trdatamat_\parentIdx$, we can compute
	\begin{equation} \label{eq:indLikelihood}
	\cdensity{\datamatobs_{\databelow{\parentIdx}}}{\trdatamat_\parentIdx, \facprecDiff, \treeparam} =  \cdensity{\datamatobs_{\databelow{\childIdxOne}}}{\trdatamat_\parentIdx, \facprecDiff, \treeparam}  \cdensity{\datamatobs_{\databelow{\childIdxTwo}}}{\trdatamat_\parentIdx, \facprecDiff, \treeparam}
	\end{equation}
	as $\datamatobs_{\databelow{\childIdxOne}}$ and $\datamatobs_{\databelow{\childIdxTwo}}$ are conditionally independent given $\trdatamat_\parentIdx$.
	Using Equations \ref{eq:diffusion_child} and \ref{eq:propMVN}, we form
	\begin{equation} \label{eq:partialMVNdensity}
	\begin{split} \cdensity{\datamatobs_{\databelow{\childIdxOne}}}{\trdatamat_\parentIdx, \facprecDiff, \treeparam} &= \int \cdensity{\datamatobs_{\databelow{\childIdxOne}}}{\trdatamat_\childIdxOne, \facprecDiff, \treeparam}\cdensity{\trdatamat_\childIdxOne}{\trdatamat_\parentIdx, \facprecDiff, \treeparam} \differential \trdatamat_\childIdxOne
	= \remainder{\childIdxOne}\degmvn{\trdatamat_\parentIdx}{\nodeMean{\childIdxOne}}{{\deflatedNodePrecision{\childIdxOne}}},
	\end{split}
	\end{equation}
	where the branch-deflated pseudo-precision $\deflatedNodePrecision{\childIdxOne} = \left(\nodePrecision{\childIdxOne}^\specialInverse + \branchLength{\childIdxOne}\notMissingMatrix{\childIdxOne}\facprecDiff\notMissingMatrix{\childIdxOne}\right)^\specialInverse$.
	See Supplementary Information (SI) Section \ref{app:bestPeel} for details on computing this pseudo-inverse.
	We use the results of Equation \ref{eq:partialMVNdensity} in Equation \ref{eq:indLikelihood} to compute the partial log-likelihood
	\begin{equation} \label{eq:dataBelowLikelihood}
	\begin{split}
	\log \cdensity{\datamatobs_{\databelow{\parentIdx}}}{\trdatamat_\parentIdx, \facprecDiff, \treeparam} &= \log  \remainder{\childIdxOne} + \log \remainder{\childIdxTwo} + \log\degmvn{\trdatamat_\parentIdx}{\nodeMean{\childIdxOne}}{{\deflatedNodePrecision{\childIdxOne}}} +
	\log\degmvn{\trdatamat_\parentIdx}{\nodeMean{\childIdxTwo}}{{\deflatedNodePrecision{\childIdxTwo}}} \\
	&= \log \remainder{\parentIdx} + \log \degmvn{\trdatamat_\parentIdx}{\nodeMean{\parentIdx}}{{\nodePrecision{\parentIdx}}},
	\end{split}
	\end{equation}
	where $\nodePrecision{\parentIdx} = \deflatedNodePrecision{\childIdxOne} + \deflatedNodePrecision{\childIdxTwo}$, $\nodeMean{\parentIdx}$ is a solution to $\nodePrecision{\parentIdx}\nodeMean{\parentIdx} = \deflatedNodePrecision{\childIdxOne}\nodeMean{\childIdxOne} + \deflatedNodePrecision{\childIdxTwo}\nodeMean{\childIdxTwo}$, and
	\begin{equation}
	\begin{split}
	\log \remainder{\parentIdx} = \log \remainder{\childIdxOne}  + \log \remainder{\childIdxTwo}  + \frac{1}{2} \log \altDet{\deflatedNodePrecision{\childIdxOne}} + \frac{1}{2} \log \altDet{\deflatedNodePrecision{\childIdxTwo}} -\frac{{\changeDimension{\childIdxOne \childIdxTwo\parentIdx}}}{2}  \log 2\pi \\
	-\frac{1}{2} \log\altDet{\nodePrecision{\parentIdx}} -\frac{1}{2}\left(\nodeMean{\childIdxOne}\transpose \deflatedNodePrecision{\childIdxOne} \nodeMean{\childIdxOne} + \nodeMean{\childIdxTwo}\transpose \deflatedNodePrecision{\childIdxTwo} \nodeMean{\childIdxTwo} - \nodeMean{\parentIdx}\transpose \nodePrecision{\parentIdx} \nodeMean{\parentIdx}\right).
	\end{split}
	\end{equation}
	Note that the change of informative dimensions $\changeDimension{\childIdxOne \childIdxTwo \parentIdx} =
	\effectiveDim{\deflatedNodePrecision{\childIdxOne}}
	+ \effectiveDim{\deflatedNodePrecision{\childIdxTwo}}
	- \effectiveDim{\nodePrecision{\parentIdx}}
	$. We update $\notMissingMatrix{\parentIdx} = \notMissingMatrix{\childIdxOne} \lor \notMissingMatrix{\childIdxTwo},$
	where $\lor$ is the element-wise ``logical or'' operation.

	Our algorithm initializes $\remainder{\taxonIdx}$, $\nodeMean{\taxonIdx}$, and $\nodePrecision{\taxonIdx}$ such that $\stochLinkobs = \remainder{\taxonIdx}\degmvn{\trdatamat_{\taxonIdx}}{\nodeMean{\taxonIdx}}{\nodePrecision{\taxonIdx}}$ at the tips of the tree.
	For the standard assumption that $\datamat_\taxonIdx = \trdatamat_\taxonIdx$, we have $\remainder{\taxonIdx} = 1$, $\nodeMean{\taxonIdx} = \nodePermutationMatrix{\taxonIdx} \left[\datamatobs_{\taxonIdx}, \mathbf{0}\right]$, and $\nodePrecision{\taxonIdx} = \nodePermutationMatrix{\nodeIndexOne}\, \diag{\raisedInfty \identityMatrix{}, 0 \identityMatrix{}} \nodePermutationMatrix{\nodeIndexOne}\transpose$.
	We perform a post-order traversal of the tree computing $\nodeMean{\parentIdx}, \nodePrecision{\parentIdx}, $ and $\remainder{\parentIdx}$ for internal nodes $\parentIdx = \ntaxa + 1, \hdots, 2\ntaxa - 2$ using the already-computed node remainders, means, and precisions for their respective child nodes.
	At the root, $\datamatobs_{\databelow{2\ntaxa -1}} = \datamatobs$ and we return the observed-data log-likelihood
	\begin{equation}\label{eq:rfull}
	\begin{split}
	\cdensity{\datamatobs}{\facprecDiff, \treeparam, \diffmean, \pss} &= \int \cdensity{\datamatobs}{\trdatamat_{2 \ntaxa - 1}, \facprecDiff, \treeparam}  \cdensity{\trdatamat_{2 \ntaxa - 1}}{\facprecDiff, \diffmean, \pss} \differential\trdatamat_{2 \ntaxa - 1} \\
	&= \int \remainder{2 \ntaxa - 1}\degmvn{\trdatamat_{2 \ntaxa - 1}}{\nodeMean{2 \ntaxa - 1}}{\nodePrecision{2 \ntaxa - 1}} \degmvn{\trdatamat_{2 \ntaxa - 1}}{\diffmean}{\pss\facprecDiff^\specialInverse} \differential\trdatamat_{2 \ntaxa - 1} \\
	&= \remainder{\full} \int\degmvn{\trdatamat_{2 \ntaxa - 1}}{\nodeMean{\full}}{\nodePrecision{\full}} \differential \trdatamat_{2 \ntaxa - 1},
	\end{split}
	\end{equation}
	where $\nodePrecision{\full} = \nodePrecision{2 \ntaxa - 1} + \pss \facprecDiff^{-1}$ and $\nodeMean{\full} = \nodePrecision{\full}\Inverse \left(\nodePrecision{2 \ntaxa - 1}\nodeMean{2 \ntaxa - 1} + \pss \facprecDiff^{-1} \diffmean\right)$.
	The integral evaluates to one, leaving the observed-data log-likelihood
	\begin{equation}
	\begin{split}
	\log\cdensity{\datamatobs}{\facprecDiff, \treeparam, \diffmean, \pss} &= \log \remainder{\full} \\
	&= \log \remainder{2 \ntaxa - 1} - \frac{\effectiveDim{\nodePrecision{2 \ntaxa - 1}}}{2} \log 2 \pi   \\
	&\hspace{1cm} + \frac{1}{2}\log \altDet{\nodePrecision{2 \ntaxa - 1}}  + \frac{1}{2} \log \altDet{\pss \facprecDiff^{-1}} - \frac{1}{2} \log \altDet{\nodePrecision{\full}} \\
	&\hspace{1cm} -\frac{1}{2}\left(\nodeMean{2 \ntaxa - 1}\transpose\nodePrecision{2 \ntaxa - 1} \nodeMean{2 \ntaxa - 1} + \pss \diffmean \transpose \facprecDiff^{-1} \diffmean -  \nodeMean{\full}\transpose\nodePrecision{\full} \nodeMean{\full}\right).
	\end{split}
	\end{equation}
	This tree traversal visits each node in $\treeparam$ exactly once and inverts a $\ntraits \times \ntraits$ matrix each time, resulting in an overall computational complexity of $\bigo{\ntaxa\ntraits^3}$ for each likelihood evaluation.

	\subsection{Inference} \label{sec:inference}
	
	The primary parameter of scientific interest is the diffusion variance $\facprecDiff$.
	We are also often interested in additional hyper-parameters $\linkParam$ related to the stochastic link function $\stochLink$.
	In cases where the tree structure is unknown, we use sequence data $\seqData$ to simultaneously infer $\treeparam$.
	As such, from a Bayesian perspective, we are interested in approximating
	\begin{equation}
	\condDensity{\facprecDiff, \treeparam, \linkParam}{\datamatobs, \seqData} \propto \condDensity{\datamatobs}{\sciParams} \density{\treeparam, \seqData} \density{\facprecDiff} \density{\linkParam},
	\label{eq:posterior}
	\end{equation}
	for inference.
	We place a $\text{Wishart}_{\ntraits}\left(\facprecDiffPrior, \df \right)$ prior on $\facprecDiff^{-1}$, where $\facprecDiffPrior$ is a $\ntraits \times \ntraits$ rate matrix.
	The prior on $\linkParam$ depends the problem of interest, and there are many ways to specify $\density{\treeparam, \seqData}$ (see \citealp{suchard2018bayesian}).
	To approximate the posterior distributions via MCMC simulation, we apply a random scan Metropolis-within-Gibbs \citep{liu1995covariance} approach by which we sample parameter blocks one at a time at random from their full conditional distribution.
	
	
	Let $\trdatamat = \left(\trdatamat_1\transpose, \cdots, \trdatamat_\ntaxa\transpose\right)\transpose$ be the latent trait values at the tips of the phylogeny.
	The conjugate $\text{Wishart}_{\ntraits}\left(\facprecDiffPrior, \df \right)$ prior on $\facprecDiff\Inverse$ implies that
	\begin{multline} \label{eq:diffPrecPosterior}
	\facprecDiff\Inverse \mid \trdatamat, \treeparam, \diffmean, \pss, \df, \facprecDiffPrior \sim \\\text{Wishart}_{\ntraits}\left[\facprecDiffPrior + \left(\trdatamat - \onevec{\ntaxa}\diffmean\transpose\right)\transpose \left(\treeVariance + \frac{1}{\pss}\oneMatrix{\ntaxa}\right)^{-1} \left(\trdatamat - \onevec{\ntaxa}\diffmean\transpose\right), \nu + N \right].
	\end{multline}
	We apply the post-order computation method proposed by \citet{ho2014linear} to compute $ \left(\trdatamat - \onevec{\ntaxa}\diffmean\transpose\right)\transpose \left(\treeVariance + \frac{1}{\pss}\mathbf{J}\right)^{-1} \left(\trdatamat - \onevec{\ntaxa}\diffmean\transpose\right)$, which has computational complexity $\bigo{\ntaxa\ntraits^2}$.
	When $\trdatamat$ are known (i.e. when there are no missing values and $\stochLink$ is degenerate at $\trdatamat_\taxonIdx$), we can sample from the distribution in Equation \ref{eq:diffPrecPosterior} immediately without any additional steps.
	However, if either assumption is violated, we must first draw from the full conditional distribution of $\trdatamat$ via the data augmentation algorithm described below.

	\subsubsection{Pre-Order Missing Data Augmentation Algorithm} \label{sec:data_augmentation}
	
	To sample jointly from the full conditional of $\trdatamat = \left(\trdatamat_1, \hdots, \trdatamat_\ntaxa\right)\transpose$, we draw on the calculations made in Section \ref{sec:diffusion_postorder} and perform a pre-order traversal of the tree.
	Note that we omit explicit dependence on the parameters $\facprecDiff, \treeparam$, and $\linkParam$ in all calculations below for clarity.
	Starting at the root, $\trdatamat_\nroot$, we draw from $\cdist{\trdatamat_\nroot}{\datamatobs, \diffmean, \pss}$.
	Using Bayes' rule and Equation \ref{eq:rfull}, we see that
	\begin{equation}
	\begin{split}
	\cdensity{\trdatamat_\nroot}{\datamatobs, \diffmean, \pss} &\propto \cdensity{\datamatobs}{\trdatamat_\nroot}\cdensity{\trdatamat_\nroot}{\diffmean, \pss}\\
	&\propto \degmvn{\trdatamat_\nroot}{\nodeMean{\full}}{\nodePrecision{\full}}\text{, which implies that}\\
	\cdist{\trdatamat_\nroot}{\datamatobs, \diffmean, \pss} &\sim \mvnDensity{\nodeMean{\full}}{\nodePrecision{\full}}.
	\end{split}
	\end{equation}
	After sampling the root traits from their full conditional, we continue the traversal of the tree where we sample each node $\trdatamat_\nodeIdxPre$ conditional on its (previously sampled) parent node $\trdatamat_{\parent{\nodeIdxPre}}$ and the observed data below node $\nodeIdxPre$ $\datamatobs_{\databelow{\nodeIdxPre}}$ for $\nodeIdxPre = 1, \hdots, 2\ntaxa - 2$.
	For the internal nodes, we compute $\cdensity{\trdatamat_\nodeIdxPre}{\datamatobs_{\databelow{\nodeIdxPre}}, \trdatamat_{\parent{\nodeIdxPre}}}$ as follows:
	\begin{equation}\label{eq:preInternal}
	\begin{split}
	\cdensity{\trdatamat_\nodeIdxPre}{\datamatobs_{\databelow{\nodeIdxPre}}, \trdatamat_{\parent{\nodeIdxPre}}} &\propto \cdensity{\datamatobs_{\databelow{\nodeIdxPre}}}{\trdatamat_\nodeIdxPre}\cdensity{\trdatamat_\nodeIdxPre}{\trdatamat_{\parent{\nodeIdxPre}}}\\
	&\propto \degmvn{\trdatamat_\nodeIdxPre}{\nodeMean{\nodeIdxPre}}{\nodePrecision{\nodeIdxPre}} \degmvn{\trdatamat_\nodeIdxPre}{\trdatamat_{\parent{\nodeIdxPre}}}{\left(\branchLength{\nodeIdxPre}\facprecDiff\right)\Inverse}\\
	&\propto \degmvn{\trdatamat_\nodeIdxPre}{\simMean{\nodeIdxPre}}{\simPrec{\nodeIdxPre}}
	\end{split}
	\end{equation}
	where $\simPrec{\nodeIdxPre} = \nodePrecision{\nodeIdxPre} + \left(\branchLength{\nodeIdxPre}\facprecDiff\right)\Inverse$ and $\simMean{\nodeIdxPre} = \simPrec{\nodeIdxPre}\Inverse\left(\nodePrecision{\nodeIdxPre}\nodeMean{\nodeIdxPre} + \left(\branchLength{\nodeIdxPre}\facprecDiff\right)\Inverse \trdatamat_{\parent{\nodeIdxPre}}\right)$.
	This implies $\cdist{\trdatamat_\nodeIdxPre}{\datamatobs_{\databelow{\nodeIdxPre}}, \trdatamat_{\parent{\nodeIdxPre}}} \sim \mvnDensity{\simMean{\nodeIdxPre}}{\simPrec{\nodeIdxPre}}$, and we sample $\trdatamat_\nodeIdxPre$ from this distribution.
	
	At the tips, we employ one of two techniques depending on the specific model. Under our standard assumption (i.e. $\trdatamat_\taxonIdx = \datamat_\taxonIdx$ with probability 1), we partition the precision $\facprecDiff\Inverse$ and trait values $\trdatamat_\taxonIdx$ and $\trdatamat_{\parent{\taxonIdx}}$ such that
	
	\begin{equation}
	\facprecDiff\Inverse = \permutationMatrix_\taxonIdx \left(\begin{matrix} \dooBlock{\taxonIdx} & \domBlock{\taxonIdx} \\ \dmoBlock{\taxonIdx} & \dmmBlock{\taxonIdx} \end{matrix}\right) \permutationMatrix_\taxonIdx\transpose,
	\quad \trdatamat_\taxonIdx = \permutationMatrix_\taxonIdx \left(\begin{matrix}\trdatamatobs_\taxonIdx \\\trdatamatmis_\taxonIdx \end{matrix}\right),
	\text{ and}\quad \trdatamat_{\parent{\taxonIdx}} = \permutationMatrix_\taxonIdx \left(\begin{matrix}\trdatamatobs_{\parent{\taxonIdx}} \\\trdatamatmis_{\parent{\taxonIdx}} \end{matrix}\right)
	\end{equation}
	and draw from $\cdist{\trdatamatmis_\taxonIdx}{\datamatobs_\taxonIdx, \trdatamat_{\parent{\taxonIdx}}} \sim \mvnDensity{\trdatamatmis_{\parent{\taxonIdx}} + {\dmmBlock{\taxonIdx}}\Inverse \dmoBlock{\taxonIdx}\left(\trdatamatobs_{\parent{\taxonIdx}} - \trdatamatobs_\taxonIdx\right)}{\frac{1}{\branchLength{\taxonIdx}}\dmmBlock{\taxonIdx}}$ for $\taxonIdx = 1, \hdots, \ntaxa$.
	For cases where $\stochLink$ is non-degenerate, we simply use Equation \ref{eq:preInternal} to sample from $\cdist{\trdatamat_\taxonIdx}{\datamatobs_\taxonIdx, \trdatamat_{\parent{\taxonIdx}}}$.
	Once we have sampled $\condDist{\trdatamat}{\datamatobs, \facprecDiff, \treeparam, \linkParam}$, we can draw from the full conditional distribution of $\facprecDiff\Inverse$ via Equation \ref{eq:diffPrecPosterior}.
	This pre-order data augmentation procedure requires a single $\ntraits\times\ntraits$ matrix inversion at each of the $\nroot$ nodes in the tree, resulting in overall computational complexity $\bigo{\ntaxa\ntraits^3}$.

	\section{Model Extension: Residual Variance}
	\label{sec:sampling_error}

	We extend the MBD model of phenotypic evolution to include multivariate normal residual variance at each of the tips.
	Under this model, we assume
	\begin{equation} \label{eq:samplingModel}
	\condDensity{\datamat_\taxonIdx}{\trdatamat_\taxonIdx} = \degmvn{\datamat_\taxonIdx}{\trdatamat_\taxonIdx}{\samplingPrec} \text{ for } \taxonIdx = 1, \ldots, \ntaxa ,
	\end{equation}
	where $\samplingPrec$ is a $\ntraits \times \ntraits$ precision matrix.
	Under this model, the vectorization of $\datamat$ is MVN-distributed with $\ntaxa\ntraits\times\ntaxa\ntraits$ variance-covariance matrix $\variance \kronecker \left( \treeAndRootVariance \right) + \samplingVar \kronecker \identityMatrix{\ntaxa}$ where $\identityMatrix{\ntaxa}$ is the $\ntaxa \times \ntaxa$ identity matrix.
	Unlike the case where $\datamat_\taxonIdx = \trdatamat_\taxonIdx$,  $\datamat$ cannot be expressed as matrix-normal even in the data-complete case because the variance cannot be expressed as the Kronecker product of two matrices. As such, our post-order likelihood computation algorithm is useful for this extended model, even when there are no missing data points.
	
	\subsection{Inference of Residual Variance} \label{sec:sampling_inference}
	
	Similar to our inference of $\facprecDiff$ in the diffusion process, we place a conjugate $\text{Wishart}\left(\samplingScale, \samplingDf\right)$ prior on $\samplingPrec$ using the rate parameterization.
	This yields the full conditional distribution
	\begin{equation} \label{eq:samplingPosterior}
	\condDist{\samplingPrec}{\datamat, \trdatamat} \sim \text{Wishart}_\ntraits\left(\samplingScale + \left(\datamat - \trdatamat\right)\transpose \left(\datamat - \trdatamat\right), \samplingDf + \ntaxa \right).
	\end{equation}
	Because $\trdatamat$ is latent in this model, each time we update $\samplingPrec$ we first draw from the full conditional posterior of $\trdatamat$ using the algorithm described in Section \ref{sec:data_augmentation}.
	For cases where $\datamat$ is not completely observed, we must perform an additional data augmentation step where we draw from $\condDist{\datamatmis}{\datamatobs, \trdatamat, \samplingPrec}$.
	To do this, we decompose the sampling precision matrix into blocks such that
	\begin{equation}
	\samplingPrec = \permutationMatrix_\taxonIdx \left(\begin{matrix} \pooBlock{\taxonIdx} & {\pmoBlock{\taxonIdx}}\transpose \\ \pmoBlock{\taxonIdx} & \pmmBlock{\taxonIdx} \end{matrix}\right) \permutationMatrix_\taxonIdx\transpose \text{ for } i = 1, \hdots, \ntaxa.
	\end{equation}
	From Equation \ref{eq:samplingModel}, we see that
	\begin{equation} \label{eq:missingSim}
	\condDensity{\datamatmis_\taxonIdx}{\datamatobs_\taxonIdx, \trdatamat_\taxonIdx, \samplingPrec} = \degmvn{\datamatmis_\taxonIdx}{\trdatamatmis_\taxonIdx + {\pmmBlock{\taxonIdx}}\Inverse \pmoBlock{\taxonIdx}\left(\trdatamatobs_\taxonIdx - \datamatobs_\taxonIdx\right)}{\pmmBlock{\taxonIdx}}.
	\end{equation}
	As such, we can directly sample $\datamatmis_\taxonIdx$ from its full conditional above and update $\datamat_\taxonIdx = \permutationMatrix_\taxonIdx \left[\datamatobs_{\taxonIdx}, \datamatmis_\taxonIdx\right]\transpose$ for $\taxonIdx = 1, \hdots, \ntaxa$.
	This process also has computational complexity $\bigo{\ntaxa \ntraits^3}$.
	
	Note that we can draw from the joint full conditional of $\facprecDiff$ and $\samplingPrec$ by performing a single pre-order data augmentation where we draw from $\cdensity{\trdatamat, \datamatmis}{\facprecDiff, \samplingPrec}$ and subsequently draw from $\cdensity{\facprecDiff, \samplingPrec}{\trdatamat, \datamatmis} = \cdensity{\facprecDiff}{\trdatamat}\cdensity{\samplingPrec}{\trdatamat, \datamat}$.
	These distributions are conditionally independent due to the fact that $\trdatamat$ and $\trdatamat - \datamat$ are independent by construction.
	This procedure effectively halves the computation time as we only need to perform a single post-order likelihood computation/pre-order data augmentation step to sample both $\facprecDiff$ and $\samplingPrec$, rather than each time we sample one.

	\subsection{Heritability Statistic}\label{sec:heritability}

	The residual variance extension enables us to estimate phenotypic heritability over evolutionary time.
	We use a definition analogous to the broad-sense heritability in statistical genetics (see \citealp{visscher2008heritability}).
	Namely, we seek to quantify the proportion of variance in a trait attributable to the Brownian diffusion process on the phylogeny (as opposed to the residual variance).
	Note that we are primarily interested in heritability in the HIV example below, for which we use data from a recent paper by \cite{blanquart17Heritability}.
	As such, we use a multivariate generalization of the heritability statistic from that paper.
	Specifically, we estimate phylogenetic heritability by taking the expectation of the empirical sample variance under our extended model.
	We define the $\ntraits \times \ntraits$ empirical covariance matrix as
	\begin{equation}
	\begin{split}
	\sampleCovariance &= \frac{1}{\ntaxa} \sum_{\taxonIdx = 1}^{\ntaxa} \left(\datamat_\taxonIdx - \vecMean\right)\left(\datamat_\taxonIdx - \vecMean\right)\transpose = \frac{1}{\ntaxa} \left(\datamat - \matMean\right)\transpose \left(\datamat - \matMean\right),
	\end{split}
	\end{equation}
	where $\vecMean = \frac{1}{\ntaxa} \sum_{\taxonIdx = 1}^\ntaxa \datamat_\taxonIdx = \frac{1}{\ntaxa} \datamat\transpose \oneVector{\ntaxa}$ and $\matMean = \oneVector{\ntaxa}\vecMean\transpose = \frac{1}{\ntaxa}\oneMatrix{\ntaxa}\datamat$.
	The expectation of this quantity reduces to the following expression (see \siSecName~\ref{app:heritability} for details):
	\begin{equation} \label{eq:expected_covariance}
	\expectedTotalVariance = \frac{\ntaxa - 1}{\ntaxa} \samplingVar  + \left(\frac{1}{\ntaxa} \trOperator{\treeVariance} - \frac{1}{{\ntaxa}^2}\oneVector{\ntaxa}\transpose \treeVariance  \oneVector{\ntaxa}\right)\facprecDiff.
	\end{equation}
	
	Because $\expectedTotalVariance$ is a linear combination of $\facprecDiff$ and $\samplingVar$, we propose the $\ntraits \times \ntraits$ heritability matrix $\heritabilityMatrix = \{\heritabilityMatrixsm_{\traitIdxTwo \traitIdxThree}\}$ with entries
	\begin{equation} \label{eq:CoheritabilityStat}
	\heritabilityMatrixsm_{\traitIdxTwo \traitIdxThree} = \frac{\diffConstant\facprecDiffsm_{\traitIdxTwo \traitIdxThree}}{\sqrt{\left(\diffConstant\facprecDiffsm_{\traitIdxTwo \traitIdxTwo} + \resConstant\samplingVarsm_{\traitIdxTwo \traitIdxTwo}\right)\left(\diffConstant\facprecDiffsm_{\traitIdxThree \traitIdxThree} + \resConstant\samplingVarsm_{\traitIdxThree \traitIdxThree}\right)}},
	\end{equation}
	where $\diffConstant = \frac{1}{\ntaxa} \trOperator{\treeVariance} - \frac{1}{{\ntaxa}^2}\oneVector{\ntaxa}\transpose \treeVariance  \oneVector{\ntaxa}$ and $\resConstant = \frac{\ntaxa - 1}{\ntaxa}$.
	The diagonal entries $\heritabilityMatrixsm_{\traitIdxTwo\traitIdxTwo} = \heritabilityMatrixsm^2_{\traitIdxTwo}$ represent the marginal phylogenetic heritability of each trait, and the off-diagonal entries represent pair-wise co-heritability \citep[][chap.~19]{falconer1960introduction} between traits.
	Note that naive computation of $\diffConstant$ relies on constructing $\treeVariance$ that has complexity $\bigo{\ntaxa^2}$.
	\siSecName~\ref{app:treeSum} describes a novel post-order algorithm that computes $\diffConstant$ in $\bigo{\ntaxa}$ time without explicitly constructing $\treeVariance$.
	While the breadth of research in heritability is extensive across both statistical genetics and phylogenetics (see in particular the recent paper by \citealp{Mitov2018}), we choose the same heritability statistic as used by \cite{blanquart17Heritability} for direct comparison with their analysis.
	That being said, our methods could be readily adapted to approximate the posterior distribution of several of the alternative heritability statistics presented in \cite{Mitov2018}.
	Additionally, our pre-order data augmentation procedure allows us to generate samples directly from the posterior of the latent trip traits $\trdatamat$, from which we can directly compute the genetic covariance $\sampleTrCov$ rather than relying on	 expectations.

	\section{Research Materials}
	We have implemented these methods in the development version of BEAST \citep{BEAST}.
	The XML files for all analyses and instructions for running them are available at \url{https://github.com/suchard-group/missing_traits_paper}.
	\section{Computational Efficiency}\label{sec:timing}
	Our method dramatically increases computational efficiency over the current best-practice method.
	This latter procedure,  developed by \cite{cybis2015assessing}, treats the missing and latent values of $\trdatamat$ as unknown parameters and numerically integrates them out by placing a Gibbs sampler on each tip $\trdatamat_\taxonIdx$ that draws from its full conditional distribution $\condDensity{\trdatamat_\taxonIdx}{\datamat_\taxonIdx, \trdatamat_{\dataabove{\taxonIdx}}}$ for $\taxonIdx = 1, \hdots, \ntaxa$ where $\trdatamat_{\dataabove{\taxonIdx}} = \trdatamat \backslash \trdatamat_\taxonIdx$.
	Because the full conditional distribution of $\trdatamat_\taxonIdx$ relies on the other missing and latent values in $\trdatamat$, we sample each tip individually.
	The advantage of this is that the likelihood calculation, the Gibbs sampler of the diffusion variance $\facprecDiff$, and the data augmentation procedure for each tip all have complexity $\bigo{\ntaxa\ntraits^2}$ rather than our $\bigo{\ntaxa\ntraits^3}$.
	As such, this numerical integration procedure has overall complexity $\bigo{\nmissing\ntaxa\ntraits^2}$ where $\nmissing$ is the number of tips with missing or latent values.
	For any extended model where $\stochLink$ is not degenerate at $\trdatamat_\taxonIdx$, all values of $\trdatamat$ are latent and $\nmissing = \ntaxa$.
	
	We formalize our comparison by computing the median and minimum effective sample size (ESS) per hour for all parameters of interest under both our analytical integration method and the sampling method discussed above.
	We also compute the ESS per sample and samples per hour to understand how our improved method influences both the autocorrelation between MCMC samples and the amount of computation work required to generate a single draw from the posterior.
	We define the number of samples as the number of states in which the MCMC simulation updates the parameters of interest (as opposed to missing trait values).
	Table \ref{tb:timing} presents the results of our efficiency comparisons.
	
	\begin{table}
		\caption{Algorithmic improvement. We report MCMC sampling efficiency through effective sample size (ESS) that shows both a decrease in autocorrelation (as shows by ESS / Sample) and in the overall work required per sample (as shown by Samples / Hour).}
		\begin{center}
			\def\Increase{Speed-up}
			\begin{tabular}{cll rr c rr r}
				\toprule
				\multirow{2}{*}{\makecell[c]{Data\\set}} & \multicolumn{1}{c}{\multirow{2}{*}{\makecell[c]{Model}}} &\multirow{2}{*}{\makecell[c]{Integration\\method}}& \multicolumn{2}{c}{ESS/hour} && \multicolumn{2}{c}{ESS/sample} & \multirow{2}{*}{\makecell[c]{Samples\\/hour}} \\
				\cmidrule{4-5} \cmidrule{7-8}
				& & & minimum & median && minimum & median & \\
				\midrule
				\multirow{6}{*}{\rotatebox{90}{Mammals}} & \multirow{3}{*}{\makecell[l]{Diffusion\\only}}&Analytic & \minEssPHrAnMam & \medEssPHrAnMam && \minESSPStAnMam & \medESSPStAnMam & \minStPHrAnMam \\
				& & Sampling & \minEssPHrSamMam & \medEssPHrSamMam && \minESSPStSamMam & \medESSPStSamMam & \minStPHrSamMam\\
				& & \multicolumn{1}{c}{\textbf{\Increase}} & \textbf{\minEssPHrMamFold}$\times$ & \textbf{\medEssPHrMamFold}$\times$ && \textbf{\minESSPStMamFold}$\times$ & \textbf{\medESSPStMamFold}$\times$ & \textbf{\medStPHrMamFold}$\times$\\
				\cmidrule{3-9}
				& \multirow{3}{*}{\makecell[l]{Diffusion\\with residual}} & Analytic& \minEssPHrAnMamRM & \medEssPHrAnMamRM && \minESSPStAnMamRM & \medESSPStAnMamRM & \minStPHrAnMamRM \\
				& & Sampling & \minEssPHrSamMamRM & \medEssPHrSamMamRM && \minESSPStSamMamRM & \medESSPStSamMamRM & \minStPHrSamMamRM\\
				& & \multicolumn{1}{c}{\textbf{\Increase}} & \textbf{\minEssPHrMamRMFold}$\times$ & \textbf{\medEssPHrMamRMFold}$\times$ && \textbf{\minESSPStMamRMFold}$\times$ & \textbf{\medESSPStMamRMFold}$\times$ & \textbf{\medStPHrMamRMFold}$\times$ \\
				\cmidrule{2 - 9}
				\multirow{6}{*}{\rotatebox{90}{HIV}}& \multirow{3}{*}{\makecell[l]{Diffusion\\only}} & Analytic & \minEssPHrAnHIV & \medEssPHrAnHIV && \minESSPStAnHIV & \medESSPStAnHIV & \minStPHrAnHIV \\
				& & Sampling & \minEssPHrSamHIV & \medEssPHrSamHIV && \minESSPStSamHIV & \medESSPStSamHIV & \minStPHrSamHIV\\
				& & \multicolumn{1}{c}{\textbf{\Increase}} & \textbf{\minEssPHrHIVFold}$\times$ & \textbf{\medEssPHrHIVFold}$\times$ && \textbf{\minESSPStHIVFold}$\times$ & \textbf{\medESSPStHIVFold}$\times$ & \textbf{\medStPHrHIVFold}$\times$\\
				\cmidrule{3-9}
				& \multirow{3}{*}{\makecell[l]{Diffusion\\with residual}} & Analytic& \minEssPHrAnHIVRM & \medEssPHrAnHIVRM && \minESSPStAnHIVRM & \medESSPStAnHIVRM & \minStPHrAnHIVRM \\
				& & Sampling & \minEssPHrSamHIVRM & \medEssPHrSamHIVRM && \minESSPStSamHIVRM & \medESSPStSamHIVRM & \minStPHrSamHIVRM\\
				& & \multicolumn{1}{c}{\textbf{\Increase}} & \textbf{\minEssPHrHIVRMFold}$\times$ & \textbf{\medEssPHrHIVRMFold}$\times$ && \textbf{\minESSPStHIVRMFold}$\times$ & \textbf{\medESSPStHIVRMFold}$\times$ & \textbf{\medStPHrHIVRMFold}$\times$ \\
				\bottomrule
			\end{tabular}
		\end{center}
		\label{tb:timing}
	\end{table}

	\section{Applications}
	\subsection{Mammalian Life History}
	A major task for life history theory is to understand the ecological and evolutionary significance of correlation between life history traits such as age at sexual maturity, the number of offspring per reproductive event, and reproductive lifespan \citep{Roff2002}.
	Establishing patterns of such correlation grants insight into whether life history variation between individuals, populations or species is consistent with pace-of-life theory \citep{Reynolds2003, Reale2010}. This theory predicts that ‘fast’ traits such as early maturity, large broods, small offspring, frequent reproduction and a short lifespan are positively associated with each other as a consequence of organisms pursuing strategies that prioritize either current or future reproduction.
	Existing approaches using comparative life history data to investigate fast-slow trait covariation patterns (e.g. mammals: \citealp{Bielby2007}; hymenoptera: \citealp{Blackburn1991}; lizards: \citealp{Clobert1998}; birds: \citealp{Saether2000}; plants: \citealp{Salguero2017}; fish: \citealp{Wiedmann2014}) generally support the fast-slow hypothesis; however, results are rarely consistent across taxa. This may reflect important taxonomic differences in life history evolution, but there is concern that differences are an artifact of different methodologies \citep{Jeschke2009}.
	
	One key limitation is that previous methods have required complete data for each species. As complete measurements across a rich suite of varied life history traits are not yet available for most species, this means that researchers must choose to either reduce the number of traits or reduce the number of species included in analyses.
	By integrating out missing traits, we resolve this issue and analyze the life history dataset used in \cite{Capellini2015}, which is based largely on the final PanTHERIA dataset \citep{Jones2009}, supplemented with measurements from \cite{Ernest2003} and additional sources.
	Our analysis includes all the variables analyzed by \cite{Bielby2007} (gestation length, weaning age, neonatal body mass, litter size, litter frequency, and age at first birth) plus reproductive lifespan (maximum lifespan minus age at first birth).
	We include female body mass as a trait rather than analyze size-corrected residuals and log-transform and standardize all traits prior to analysis. 
	The analysis assumes the phylogeny of \cite{Fritz2009} that remains the most complete phylogeny for mammals.
	In total, 3690 species in the phylogeny have measurement of at least one trait and are included.
	Table \ref{tb:missing} reports the number of species with measurements for each trait.
	Only 518 species have complete data on all 8 traits; thus the ability to include species with partially missing traits enables inclusion of 208\% more measurements.

	%
	%
	

	\begin{table}[h]
		\caption{Missing data summary for all three examples.}
		\begin{center}
			\begin{tabular}{l l r r}
				\toprule
				Data set & Trait & \makecell[l]{Number\\observed} & \makecell[l]{Percent\\missing}\\
				\midrule
				\multirow{9}{*}{\rotatebox[]{90}{\makecell{Mammals\\\small{$\ntaxa=3690$}}}} & Body mass & 3508 & 4.9\% \\
				&Litter size & 2538 & 31.2\% \\
				&Gestation length & 1427 & 61.3\% \\
				&Weaning age	& 1253 & 66.0\% \\
				&Litter frequency & 	1231 & 66.6\% \\
				&Neonatal body mass & 1108 & 70.0\% \\
				&Age at first birth & 945 & 74.4\% \\
				&Reproductive lifespan & 748 & 79.7\% \\
				&\multicolumn{1}{c}{\textbf{Total}} &\textbf{12758} & \textbf{56.8\%} \\
				\midrule[0.5pt]
				\multirow{8}{*}{\rotatebox[]{90}{\makecell{Prokaryotes\\\small{$\ntaxa = 705$}}}} & Cell diameter & 690 & 2.1\% \\
				&Cell length & 657 & 6.8 \% \\
				&Genome length & 563 & 20.1\% \\
				&GC content & 563 & 20.1\% \\
				&Coding sequence length & 558 & 20.9\%\\
				&Optimal temperature & 548 & 22.2\% \\
				&Optimal pH & 487 & 30.9\% \\
				&\multicolumn{1}{c}{\textbf{Total}} &\textbf{4066} & \textbf{17.6\%} \\
				\midrule[0.5pt]
				\multirow{4}{*}{\rotatebox[]{90}{\makecell{HIV\\\small{$\ntaxa = 1536$}}}} & GSVL & 1536 & 0.0\% \\
				&SPVL & 1536 & 0.0\% \\
				&CD4 slope & 1102 & 28.3\% \\
				&\multicolumn{1}{c}{\textbf{Total}} &\textbf{4174} & \textbf{9.4\%} \\
				\bottomrule
			\end{tabular}
		\end{center}
		\label{tb:missing}
	\end{table}

	
	To estimate the correlation between these traits throughout mammalian evolution, we jointly model them with an MBD process on the tree with residual variance.
	In this analysis, we are primarily interested in the correlation between traits during the MBD process on the tree and estimate trait correlations from the marginal posterior of $\facprecDiff$.
	Figure \ref{fig:mammalsDiffusion} summarizes these findings.
	\def\corrWidth{0.9\textwidth}
	\begin{figure}[h]
		\begin{center}
			\includegraphics[width=\corrWidth]{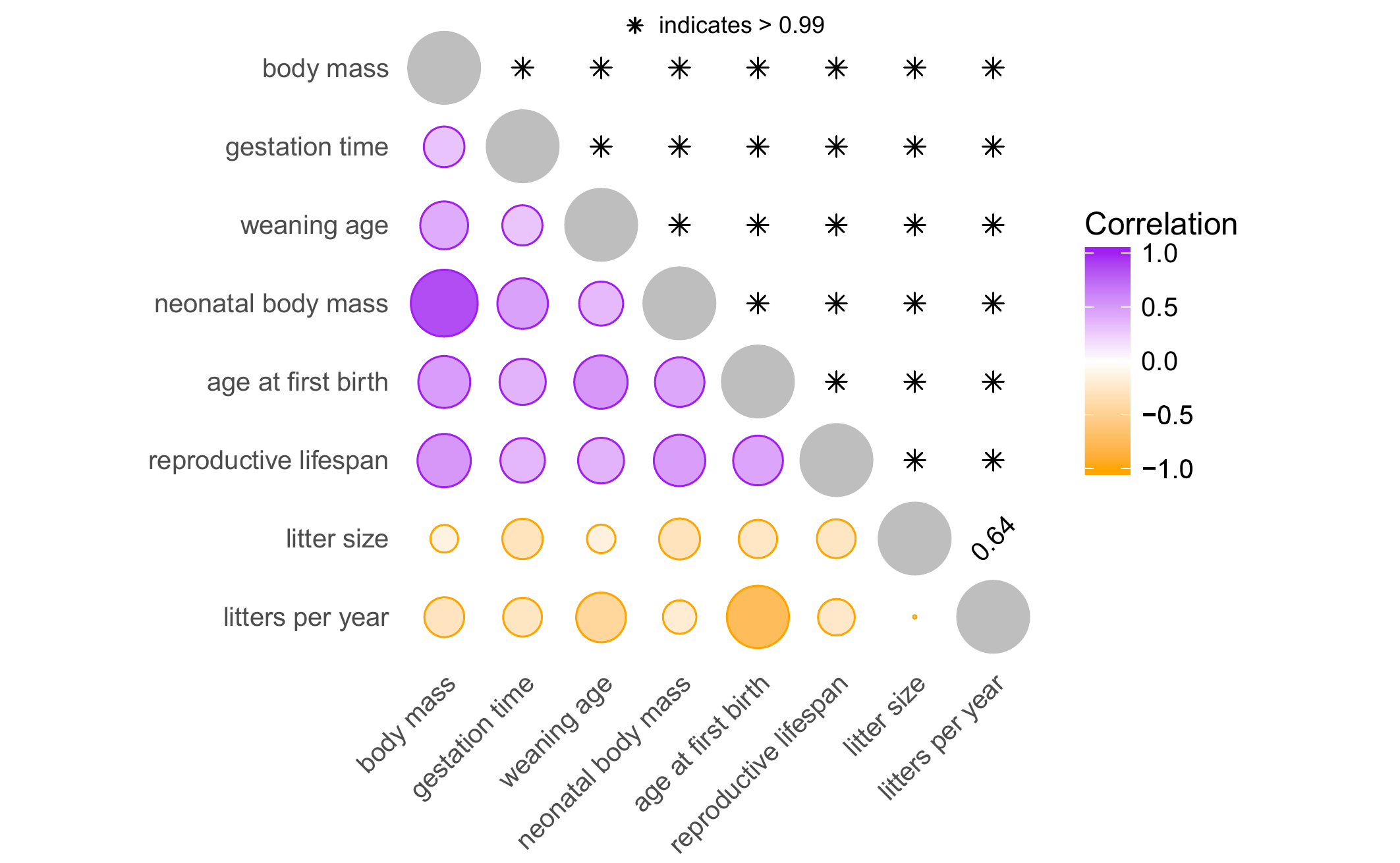}
		\end{center}
		\caption{Correlation among mammalian life-history traits. The circles below the diagonal summarize the posterior mean correlation between each pair of traits. Purple represents a positive correlation while orange represents a negative correlation. Circle size and color intensity both represent the absolute value of the correlation. The numbers above the diagonal report the posterior probability that the correlation is of the same sign as its mean.
		}
		\label{fig:mammalsDiffusion}
	\end{figure}
	Our results are clearly consistent with the fast-slow trait covariation patterns that pace-of-life theory predicts.
	The `slow' life history traits (longer gestation, later weaning, larger neonatal body mass, later age at first birth, and longer reproductive lifespan) are all positively correlated with each other and negatively correlated with the two `fast' life history traits (greater litter size and more frequent litters).
	All correlations are significant (determined by $<5\%$ posterior tail probability) with the notable exception of that between litter size and litter frequency.
	This apparent lack of correlation may be due to the opposing effects of their joint positive correlation with body mass combined with a trade-off between these two traits that life history theory predicts \citep{stearns1989}.
	Nevertheless, our results demonstrate that larger animals tend to have slower life history traits, confirming known patterns  and reflecting the central role of body size in life history evolution.
	We compare the computational efficiency of our method against that of the sampling method using the MBD model both with and without residual variance.
	Table \ref{tb:timing} shows an increase in overall computational efficiency of two orders-of-magnitude as indicated by the change in ESS per hour.
	Additionally, we see that our method succeeds at reducing both the amount of computational work per MCMC sample (as indicated by the increase in samples per hour) and autocorrelation (as indicated by the increase in ESS per sample).

	\subsection{Prokaryote evolution}
	
	Comparative genomics has greatly assisted in the formulation of prokaryote evolutionary theories.
	Several such theories have been inspired by and tested through measuring correlation among different phenotypic and genomic traits.
	For example, the thermal adaptation hypothesis posits that higher GC content is involved in adaptation to high temperatures because it may offer thermostability to genetic material \citep{bernardi1986compositional}.
	The genome streamlining hypothesis attempts to explain the compactness of prokaryotic genomes through natural selection favoring small genomes \citep{doolittle1980selfish, orgel1980selfish, giovannoni2014implications}.
	\cite{sabath2013growth} argue that lower cell volume is an adaptive response to high temperature.
	The field is well-aware of the need to account for phylogenetic relationships when measuring correlation, but statistical analyses generally rely on fixed, poorly resolved trees and simple models of trait evolution.%
	
	Here, we estimate correlation among a set of genotypic and phenotypic traits while simultaneously accounting for phylogenetic uncertainty and accommodating complexity in the trait evolutionary process.
	We construct our data set from a study by \cite{goberna2016predicting}, who collated cell diameter, cell length, optimum temperature and pH measurements for a large set of prokaryotes.
	Prior experience in resolving large, unknown trees suggests that we limit our analysis to less than $\sim$750 taxa.
	As such, we include all taxa with three or more measurements and a selection of the taxa with only two measurements in our analysis.
	For our selection of 705 taxa, we obtain data on genome length, the number of coding sequences, and GC content from the prokaryotes table in NCBI Genome.
	Table \ref{tb:missing} presents the number of measurements for each trait.
	We log-transform and standardize all traits (except for GC content which we logit-transform and standardize).
	To infer the phylogeny, we obtain matching 16S sequences via the ARB software package \citep{ludwig2004arb} that we then align using the SINA Alignment Service \citep{pruesse2012sina} and manually edit.
	
	\begin{figure}[h]
		\includegraphics[width=\corrWidth]{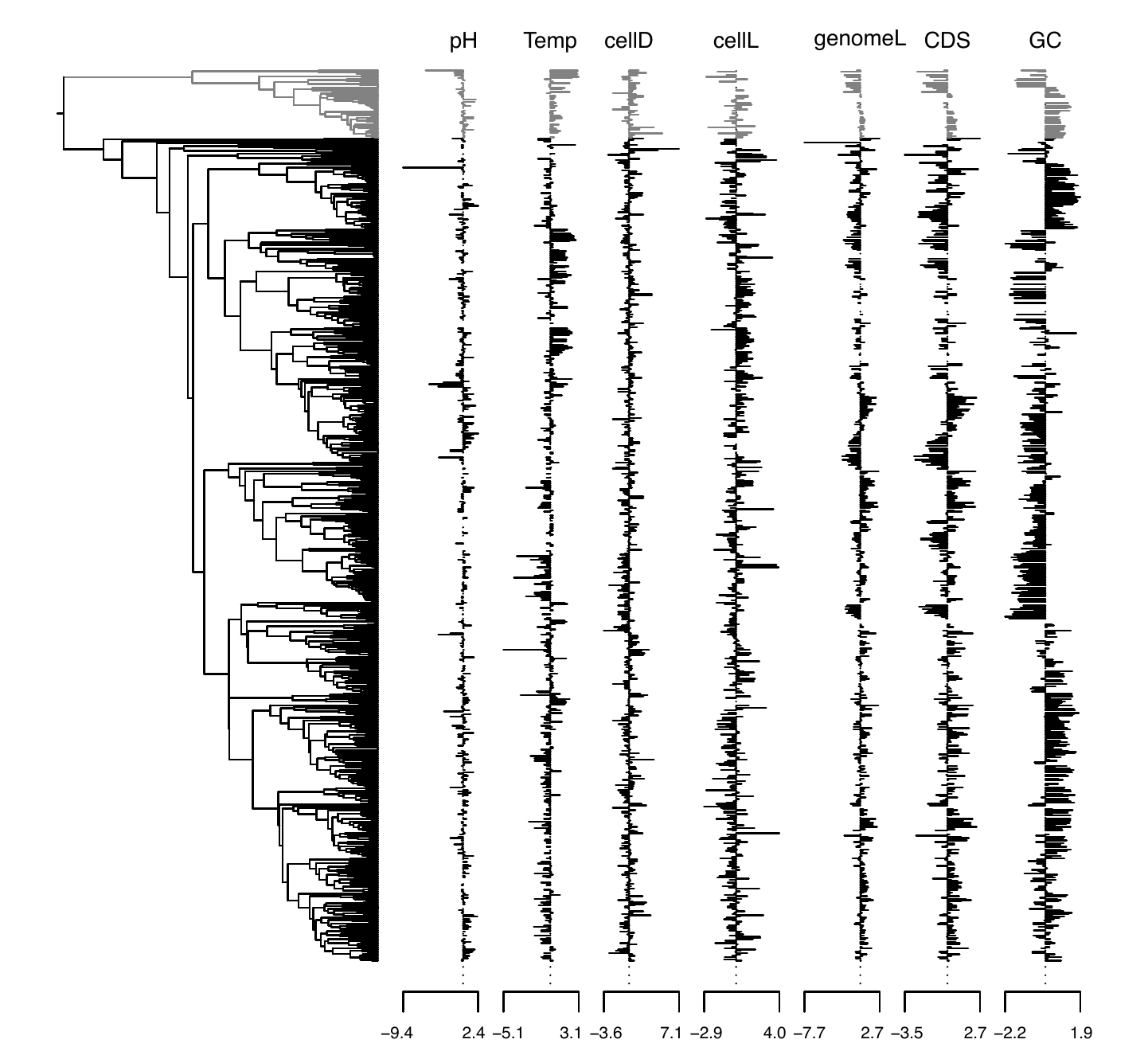}
		\caption{Prokaryote phylogeny and traits. The phylogeny depicts the inferred maximum clade credibility tree. The archaea clade ($\ntaxa=54$) and the associated trait measurements are depicted in grey.
		}
		\label{fig:prokaryoteTreeTraits}
	\end{figure}
	
	
	Through MCMC simulation, we simultaneously infer the sequence and trait evolutionary process.
	We model the sequence evolutionary process using a general time-reversible model \citep{tavare1986some} with gamma-distributed rate variation among sites \citep{yang1994maximum}. We use an uncorrelated lognormal relaxed clock to model rate variation among branches \citep{drummond2006relaxed} and specify a Yule birth prior process on the unknown tree \citep{gernhard2008conditioned}.
	For the trait evolutionary process, we assume an MBD model with residual variance.

\begin{figure}
		\begin{center}
			\includegraphics[width=\corrWidth]{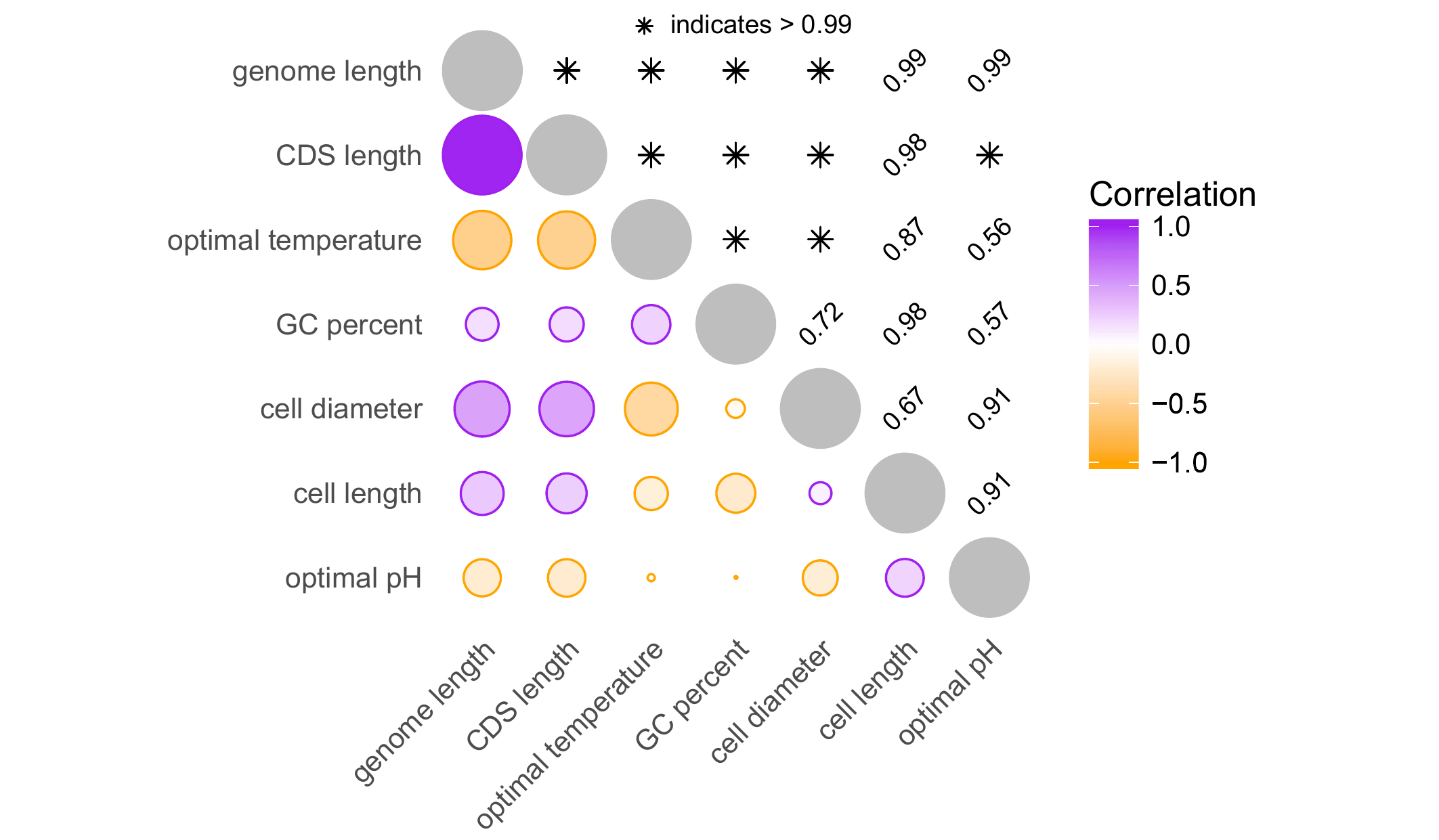}
		\end{center}
		\caption{Correlation among prokaryotic growth properties. See Figure \ref{fig:mammalsDiffusion} caption.
		}
		\label{fig:prokaryoteCorrelation}
	\end{figure}

	Figure \ref{fig:prokaryoteTreeTraits} displays our estimated maximum clade credibility phylogeny with associated trait measurements, and Figure \ref{fig:prokaryoteCorrelation}  presents the phylogenetic correlation between those traits.
	One notable result is the positive correlation between optimal temperature and GC content ($\corrOTempGCP$ posterior mean, $[\corrHDPLOTempGCP, \corrHDPHOTempGCP]$ 95\% highest posterior density interval) that the thermal adaptation hypothesis predicts \citep{bernardi1986compositional}.
	Researchers have discussed this hypothesis for years with mixed support \citep{hurst2001high, musto2004correlations, wang2006correlation, wu2012molecular, sabath2013growth, aptekmann2018core}.
	Our analysis, however, includes 435 taxa with measurements for both GC content and optimal growth temperature, making it the largest study we are aware of that accounts for phylogenetic relationships.
	Interestingly, while cell diameter and cell length are not significantly correlated, they are both positively correlated with genome length.
	Smaller cells have been associated with smaller genomes in both prokaryotes and unicellular eukaryotes \citep{shuter1983phenotypic, lynch2007frailty}, but the reasons for this are not fully understood \citep{dill2011physical}.
	We also estimate a relatively strong negative correlation between genome length and optimal temperature ($\corrOTempGenL$ $[\corrHDPLOTempGenL,\corrHDPHOTempGenL]$), supporting the genomic streamlining hypothesis during thermal adaptation.
	Note that we do not compare computation times here, as simultaneous inference of the phylogenetic tree makes the sampling method prohibitively slow.
	
	\subsection{HIV-1 virulence}
	
	Recent years have witnessed a strong interest in using phylogenetic comparative methods to study the heritability of HIV-1 virulence.
	Initially, \cite{Alizon2010} employed Pagel's $\lambda$ \citep{pagelLambda} to measure the extent to which HIV-1 set-point viral load reflects viral shared evolutionary history in the Swiss HIV Cohort Study \citep{swiss2009cohort} patients.
	A relatively high heritability estimate of set-point viral load, a predictive measure of clinical outcome, motivated others to examine to what extent the viral genotype can control for this trait (e.g. \citealp{Hodcroft2014,Vrancken2015}).
	These efforts have resulted in widely varying estimates, from 6\% to 59\%, prompting a discussion on the methods used to estimate the heritability of virulence (see \citealp{Mitov2018,Bertels2018}).
	Here, we revisit the most comprehensive data set recently analyzed \citep{blanquart17Heritability} to determine the extent to which variability in HIV-1 virulence is attributable to viral genetic variation.
	We focus on the dataset of subtype B viruses from \cite{blanquart17Heritability} that encompasses 1581 taxa with associated measures of set-point viral load and CD4 cell count decline.
	We rely on the maximum likelihood phylogeny inferred for this data set, but convert it to a time-measured tree with dated tips using a heuristic procedure \citep{To2016}.
	A prior examination of the correlation between sampling time and root-to-tip divergence using TempEst \citep{tempEst} indicated the presence of outliers, most of which can be attributed to a basal lineage in the phylogeny.
	As the subtyping of the taxa in this basal lineage also was ambiguous (Blanquart, personal communication), we remove this lineage (36 taxa) together with 9 other outlier taxa.
	We note that this resulted in a time to the most recent common ancestor (TMRCA) estimate of about 1960 that is much more in line with a recent subtype B TMRCA estimate (1967, 95\% Bayesian credibility interval of 1963–1970; \citealp{Worobey2016}) than the estimate including the basal lineage ($\sim$1930).
	
\begin{figure}[h]
		\begin{center}
			\includegraphics[width=.75\textwidth]{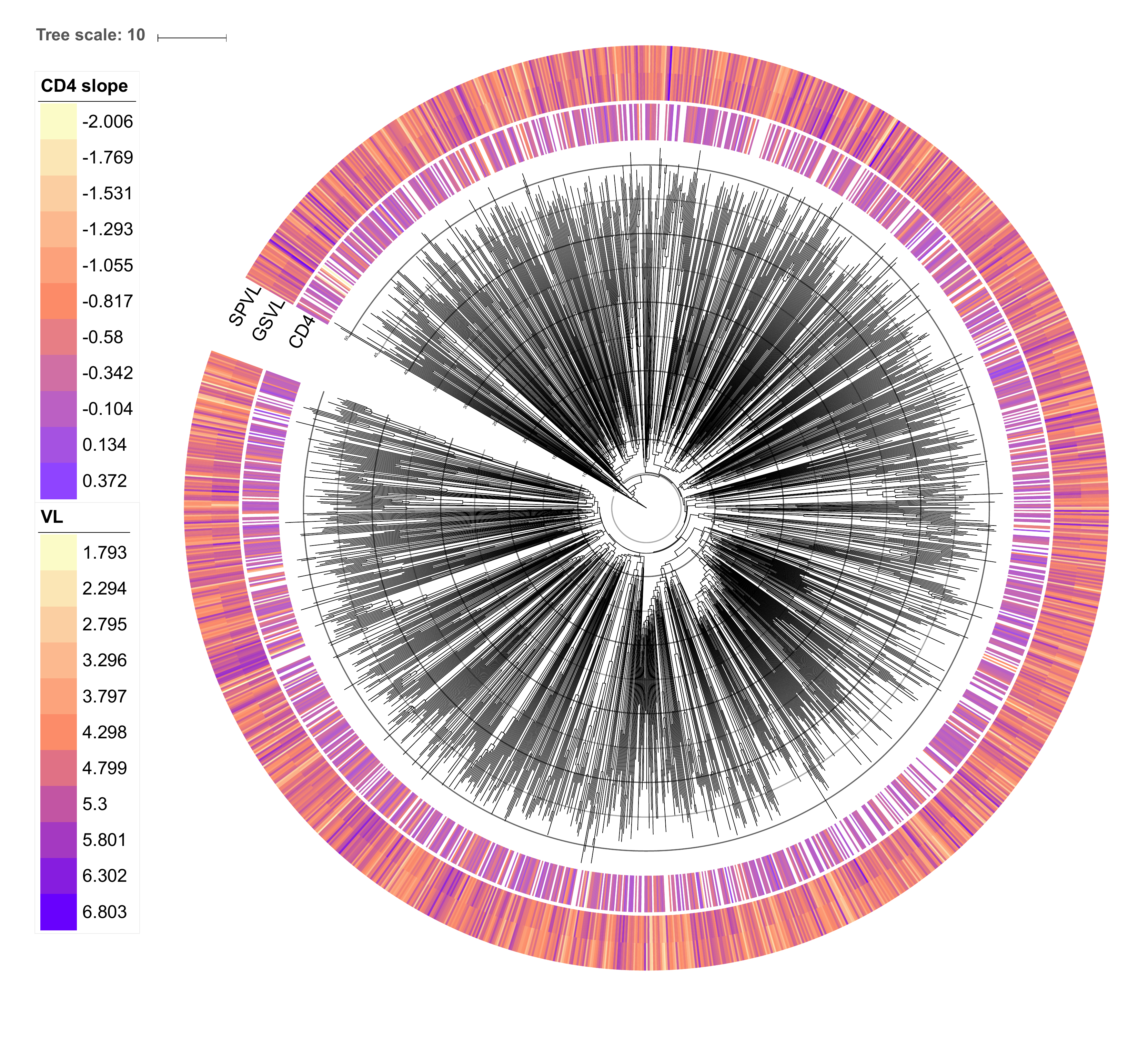}
		\end{center}
		\caption{HIV-1 phylogeny with associated CD4 slope, SPVL, and GSVL values for each viral host.}
		\label{fig:HIV_1536tre}
	\end{figure}

\begin{figure}
		\begin{center}
			\includegraphics[width=\corrWidth]{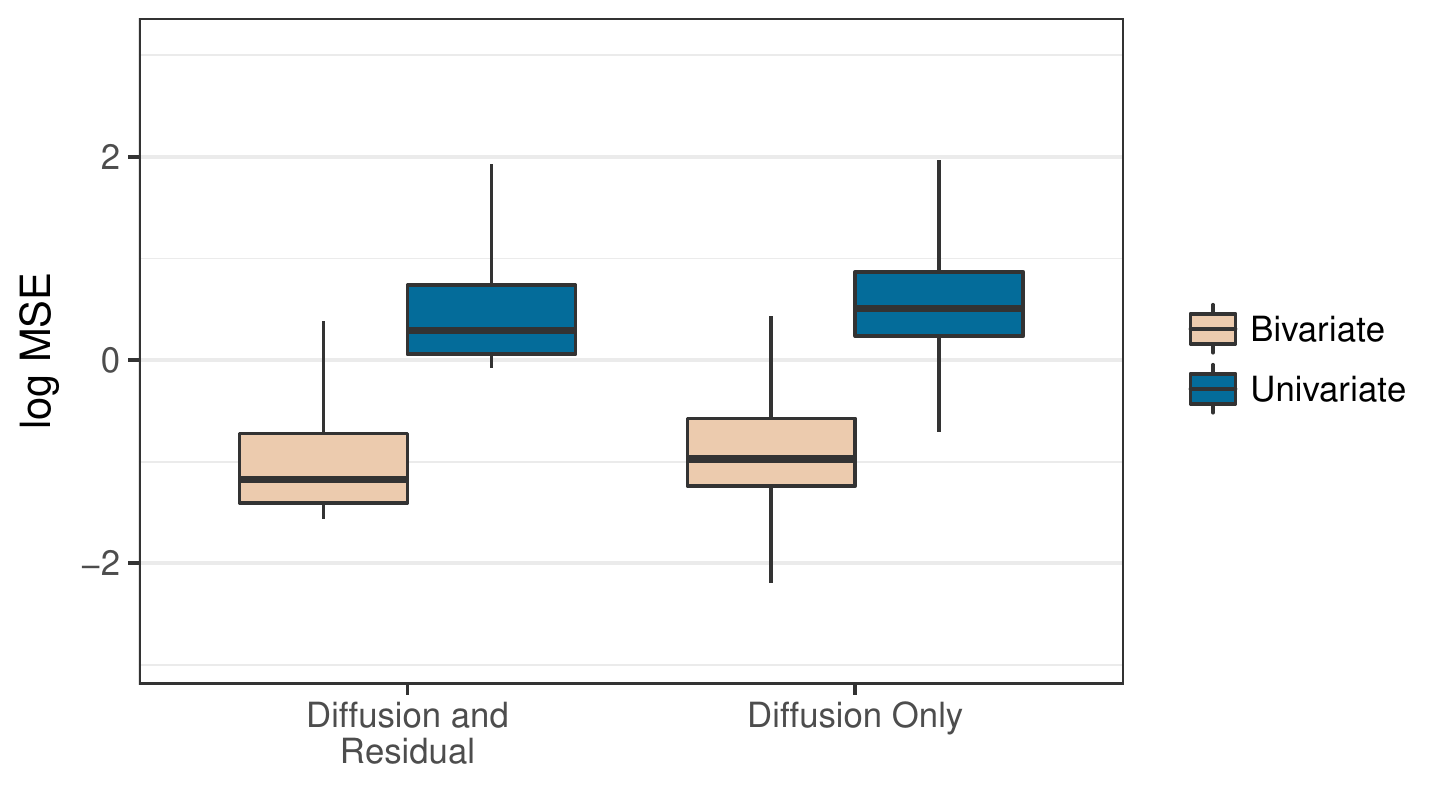}
		\end{center}
		\caption{Model predictive performance of HIV set-point viral load. Each box-and-whisker plot depicts the posterior mean-squared-error of prediction under a different model. The boxes represent the interquartile range, while the lines extend to include the $2.5^{th}$ through $97.5^{th}$ percentiles. Outliers are omitted.}
		\label{fig:HIV_MSE}
	\end{figure}

	Two measures of set-point viral load are available for all remaining taxa: (i) one based on a standardized choice of assay on a single sample taken between 6 and 24 months after infection and before the initiation of antiretroviral therapy (“gold standard viral load”, GSVL) and (ii) a more classical measure of set-point viral load (SPVL) based on the mean of all log viral loads measured between 6 and 24 months after infection.
	Figure \ref{fig:HIV_1536tre} presents the phylogeny and associated trait values.
	To estimate heritability of both set-point viral load measures and CD4 slope, we model all three measurements as a multivariate trait in our MBD model with residual variance and approximate the posterior distribution of the heritability statistic $\heritabilityMatrix$ via MCMC.
	Our estimated heritabilities are $\gsvlHeritabilityMean$ $[\gsvlHeritabilityHDPl, \gsvlHeritabilityHDPh]$ for GSVL, $\spvlHeritabilityMean$ $[\spvlHeritabilityHDPl, \spvlHeritabilityHDPh]$ for SPVL, and $\cdHeritabilityMean$ $[\cdHeritabilityHDPl, \cdHeritabilityHDPh]$
	for CD4 cell decline.
	These estimates are consistent with similar estimates reported by \cite{blanquart17Heritability}
	We further asses model fit by assessing predictive performance of GSVL on SPVL.
	We omit CD4 slope from our analysis as it is measured concurrently with SPVL.
	We randomly remove 5\% of the SPVL measurements from the data set and consider four different models.
	We consider both a bivariate case where we assume a multivariate process and a univariate case where we analyze SPVL alone.
	For both the bivariate and univariate cases, we use the MBD model both with and without the residual variance extension.
	For each removed SPVL measurement, we compute the mean squared error (MSE) between the predicted and true values.
	We repeat each analysis 50 times and report the cumulative results in Figure \ref{fig:HIV_MSE}, from which two results emerge.
	First, the MSE of prediction in the bivariate cases are lower than those in the univariate cases.
	This is unsurprising given the strong correlation between SPVL and GSVL.
	Second, addition of residual variance to the model results in modestly better prediction of SPVL in both the bivariate and univariate cases.
	This emphasizes the importance of including model extensions like residual variance in these analyses.
	
	We again demonstrate improvements in computational efficiency (see Table \ref{tb:timing}).
	While less dramatic than the mammals example, we still see an order-of-magnitude increase in effective sample size per hour in the MBD model without residual variance.
	This attenuation is to be expected, as there are far fewer missing measurements in the HIV data set than the mammal data set.
	Nevertheless, our method still outperforms the sampling method in the simple MBD model even when only 9.4\% of data points are missing.
	For the model with residual variance, our method outperforms the sampling method by two orders-of-magnitude.

	\section{Discussion}
	
	Oftentimes comparative biologists are interested in phylogenetically adjusted methods for assessing relationships between traits of organisms.
	However, frequently when the number of taxa grows large the level of missing data increases, making inference challenging.
	Here, we have developed a method for evaluating the likelihood of observed traits given a tree while integrating out missing values analytically that dramatically outperforms current best-practice methods.
	In the mammalian data set, with $\ntaxa = 3690$ and $56.8\%$ missing data, we achieve a minimum effective sample size per hour $\minEssPHrMamFold\times$ greater than previous methods.
	This increase in speed brings computation times down from more than a week to less than an hour.
	Even in the more tractable HIV data set, with $\ntaxa = 1536$ and $9.4\%$ missing data, we increase the minimum ESS per hour by a factor of $\minEssPHrHIVFold$.
	Both increases in speed are due to an overall decrease in both autocorrelation between MCMC samples and the amount of computational work required per sample.
	Importantly, this increase in computational efficiency allows for previously intractable analyses on large trees.
	Specifically, we incorporate residual variance into the model and (in the prokaryotes example) simultaneously infer $\facprecDiff$, $\samplingPrec$, and an unknown phylogeny $\treeparam$.
	Further, the residual variance extension is only one of several potential extensions.
	Other possible extensions could incorporate data sets with repeated measurements at the tips of the tree and factor analyses \citep{MaxPaper1}.

	An important limitation of our and previous methods is that they assume an ignorable missing data mechanism (i.e.~that the data are missing at random and that the prior on any model parameters is independent of the missing data mechanism) \citep{little1987statistical}.
	While these conditions may hold in some comparative biology examples, possible violations abound (e.g. any data set where values below some detection limit are omitted).
	One potential solution explicitly includes these thresholds in the model and renormalizes the observed-data likelihood appropriately.
	
	\newlength{\colshift}
	\setlength{\colshift}{-3mm}
	\begin{figure}
		\def\graphUp{2}
		\def\graphDown{-1}
		\def\graphLeft{-1.73}
		\def\graphRight{1.73}
		\def\newZero{0}

		\begin{minipage}{0.35\textwidth}
			\begin{tikzpicture}[extended line/.style={shorten >=-#1,shorten <=-#1},
			extended line/.default=1cm,
			one end extended/.style={shorten >=-#1},
			one end extended/.default=1cm,
			]
			\begin{scope}[yshift=-0cm]
			\shadedraw[ball color=black!50, opacity=0.2] (0, \newZero)  circle (0.35) node[opacity=1.0,minimum size=0.6cm,text width=2cm,align=center] (A)  {$\node_\nodeIndZero$ };
			\shadedraw[ball color=black!50, opacity=0.2] (0, \graphUp)  circle (0.35) node[opacity=1.0,minimum size=0.6cm,text width=2cm,align=center] (B)  {$\node_\nodeIndOne$ };
			\shadedraw[ball color=black!50, opacity=0.2] (\graphLeft, \graphDown)  circle (0.35) node[opacity=1.0,minimum size=0.6cm,text width=2cm,align=center] (C)  {$\node_\nodeIndTwo$ };
			\shadedraw[ball color=black!50, opacity=0.2] (\graphRight, \graphDown)  circle (0.35) node[opacity=1.0,minimum size=0.6cm,text width=2cm,align=center] (D)  {$\node_\nodeIndThree$ };
			
			\draw[extended line = -1mm] (A) -- (B) node[midway,right] {$\edge_\edgeIndOne$};
			\draw[extended line = 2mm] (A) -- (C) node[midway,above left] {$\edge_\edgeIndTwo$};
			\draw[extended line = 2mm] (A) -- (D) node[midway,above right] {$\edge_\edgeIndThree$};
			\end{scope}
			
			\end{tikzpicture}
		\end{minipage}
		\begin{minipage}{0.65\textwidth}
			\begin{tabular}{c c}
				\multicolumn{2}{c}{\textbf{Sample covariance matrices}}\\
				\toprule
				&\vspace{-2mm}\\
				$\addMat_1 = \left(\begin{matrix}\edge_\edgeIndOne &0 & 0 \\ 0 & \edge_\edgeIndTwo & 0 \\ 0 & 0 & \edge_\edgeIndThree\end{matrix}\right)$ & $\addMat_2 = \left(\begin{matrix}\edge_\edgeIndOne & \edge_\edgeIndOne & \hspace{\colshift} \edge_\edgeIndOne \\ \edge_\edgeIndOne & \edge_\edgeIndTwo + \edge_\edgeIndOne & \hspace{\colshift} \edge_\edgeIndOne \\ \edge_\edgeIndOne & \edge_\edgeIndOne & \hspace{\colshift} \edge_\edgeIndThree + \edge_\edgeIndOne\end{matrix}\right)$
			\end{tabular}
		\end{minipage}

		\caption{An acyclic graph with nodes $\{\node_\nodeIndZero, \node_\nodeIndOne, \node_\nodeIndTwo, \node_\nodeIndThree\}$ and edge weights $\{\edge_\edgeIndOne, \edge_\edgeIndTwo, \edge_\edgeIndThree \}$.
			The covariance matrix $\addMatrix = \left\{\addMatrixsm_{ij}\right\}$ is additive on an acyclic graph if each $\addMatrixsm_{ij}$ is equal to the sum of the shared non-negative edge-weights in the paths from $\node_i$ and $\node_j$ to some origin node.
			For example, the matrix $\addMat_1$ is additive for nodes $\left(\node_\nodeIndOne, \node_\nodeIndTwo, \node_\nodeIndThree\right)\transpose$ with $\node_\nodeIndZero$ at the origin, while the matrix $\addMat_2$ is additive for nodes $\left(\node_\nodeIndZero, \node_\nodeIndTwo, \node_\nodeIndThree\right)\transpose$ with $\node_\nodeIndOne$ at the origin.}
		\label{fig:acyclicGraph}
	\end{figure}

	Finally, and perhaps most importantly, we propose our method as a special case solution to the long-standing statistical problem involving multivariate normal distributions with missing data.
	Specifically, our method applies to any MVN distribution with a three-point structured covariance matrix \citep[see][]{ho2014linear}.
	Intuitively, this condition arises in covariance matrices generated from processes that are additive on an acyclic graph (see Figure \ref{fig:acyclicGraph}).
	This restriction, however, is not overly limiting and applies to a broad range of normal models including multilevel hierarchical models and matrix-normal distributions such as the one we use here.
	Additionally, our pre-order data augmentation procedure enables $\bigo{\ntaxa}$ imputation in these highly structured models.
	While \cite{allen2010transposable} and \cite{glanz2013expectation} have utilized the EM algorithm \citep{Dempster77maximumlikelihood} to efficiently perform maximum likelihood imputation in similar problems, our method could serve as an alternative for approaches that base inference on the observed-data likelihood.

	\section*{Supplementary Material}
	\begin{description}
		\item[\siName:] Includes \siSecName s 1 through 3 (pdf file)
	\end{description}

	\section*{Funding}
	
	This work was partially supported by the NIH under training programs T32-GM008185 and T32-HG002536 and Grants R01 AI107034 and U19 AI135995; NSERC Discovery under Grant RGPIN-2018-05447 and Launch Supplement DGECR-2018-00181; the Artic Network via the Wellcome Trust under project 206298/Z/17/Z; the KU Leuven Special Research Fund (`Bijzonder Onderzoeksfonds') under Grant OT/14/115; the Research Foundation -- Flanders (`Fonds voor Wetenschappelijk Onderzoek -- Vlaanderen') under Grants G066215N, G0D5117N and G0B9317N; the NSF under Grant DMS 1264153; the European Research Council via the European Union's Horizon 2020 Research and Innovation Programme under Grant no.~725422-ReservoirDOCS; and startup funds from Dalhousie University and the Canada Research Chairs Program.
	
	\section*{Acknowledgements}
	
	The authors thank Fran\c{c}ois Blanquart and Christophe Fraser for their assistance in compiling the HIV-1 data set,
	and Marta Goberna for advice in compiling the prokaryote data set.


	 \clearpage
	 
	 \setcounter{equation}{0}
	 \setcounter{section}{0}
	 \setcounter{page}{1}
	 
	 \def\suppInfo{\uppercase{SI}}
	 \titleformat{\section}{\sffamily\mdseries\upshape\uppercase}{\suppInfo\space\thesection.}{1em}{\hypertarget{myhypertarget\thesection}}
	 \titleformat{\subsection}{\sffamily\mdseries\upshape}{\suppInfo\thesubsection}{1em}{\hypertarget{myhypertarget\thesection}}
	 \titleformat{\subsubsection}{\sffamily\mdseries\upshape}{\suppInfo\thesubsubsection}{1em}{\hypertarget{myhypertarget\thesection}}
	 
	 \noindent{\Huge{\siName}}
	 
	 
	 \onehalfspacing
	 
	 \section{Matrix inversion computations}
	 \label{app:bestPeel}
	 
	 
	 
	 \def\varBlock{\diffusionVariance}
	 \newcommand{\vmmBlock}[1]{\varBlock^\text{\tiny{{mis}}}_{#1}}
	 \newcommand{\vooBlock}[1]{\varBlock^\text{\tiny{{obs}}}_{#1}}
	 \newcommand{\vomBlock}[1]{\varBlock^\text{\tiny{{om}}}_{#1}}
	 \newcommand{\vllBlock}[1]{\varBlock^\text{\tiny{{lat}}}_{#1}}
	 \newcommand{\volBlock}[1]{\varBlock^\text{\tiny{{ol}}}_{#1}}
	 \newcommand{\vlmBlock}[1]{\varBlock^\text{\tiny{{lm}}}_{#1}}
	 
	 \newcommand{\finiteObservedVariance}[1]{\volBlock{#1}}
	 \newcommand{\observedVariance}[1]{\vooBlock{#1}}
	 \newcommand{\finiteVariance}[1]{\vllBlock{#1}}
	 \def\invertibleBlock{\mathbf{T}}

	 To evaluate the observed data likelihood, we must compute branch-deflated precisions $\deflatedPrecision{\nodeIndexOne} = \left(
	 \startingPrecision{\nodeIndexOne}^{\specialInverse} + t_{\nodeIndexOne} \notMissingMatrix{\nodeIndexOne} \diffusionVariance \notMissingMatrix{\nodeIndexOne}
	 \right)^{\specialInverse}$ for $\nodeIndexOne = 1, \hdots, 2\ntaxa - 2$. We demonstrate below that this matrix exists and is well-defined under the definition of our pseudo-inverse. Using the permutation matrix $\permutationMatrix_\nodeIndexOne$ from Section \ref{sec:diffusion_postorder}, we decompose the diffusion variance $\diffusionVariance$ and node precision $\nodePrecision{\nodeIndexOne}$ such that
	 \begin{equation}\nonumber
	 \begin{aligned}
	 \diffusionVariance &= \permutationMatrix_\nodeIndexOne
	 \left( \begin{array}{ccc}
	 \vooBlock{\nodeIndexOne} & \volBlock{\nodeIndexOne} & \vomBlock{\nodeIndexOne} \\
	 - & \vllBlock{\nodeIndexOne} & \vlmBlock{\nodeIndexOne} \\
	 - & - & \vmmBlock{\nodeIndexOne}
	 \end{array} \right)
	 \permutationMatrix_\nodeIndexOne\transpose \text{ and }\\
	 \nodePrecision{\nodeIndexOne} &= \permutationMatrix_\nodeIndexOne\diag{ \raisedInfty \identityMatrix{}, \finitePrecision{\nodeIndexOne}, 0 \identityMatrix{}}\permutationMatrix_\nodeIndexOne\transpose,
	 \end{aligned}
	 \end{equation}
	 for $\nodeIndexOne = 1, \hdots, 2\ntaxa - 2$.
	 We use this decomposition to identify that:
	 \begin{equation}
	 \begin{aligned}
	 \deflatedPrecision{\nodeIndexOne} &=
	 \left(
	 \startingPrecision{\nodeIndexOne}^{\specialInverse} + t_{\nodeIndexOne} \notMissingMatrix{\nodeIndexOne} \diffusionVariance \notMissingMatrix{\nodeIndexOne}
	 \right)^{\specialInverse} \\
	 %
	 %
	 &= \nodePermutationMatrix{\nodeIndexOne}\,
	 \left(
	 \left(
	 \diag{ \raisedInfty \identityMatrix{}, \finitePrecision{\nodeIndexOne}, 0 \identityMatrix{}}
	 \right)^{\specialInverse}
	 +
	 \diag{ t_{\nodeIndexOne}
	 	\left(
	 	\begin{array}{ll}
	 	\observedVariance{\nodeIndexOne}  & \finiteObservedVariance{\nodeIndexOne} \\
	 	- & \finiteVariance{\nodeIndexOne} \\
	 	\end{array}
	 	\right), 0 \identityMatrix{}
	 }
	 \right)^{\specialInverse}
	 \nodePermutationMatrix{\nodeIndexOne}\transpose \\
	 %
	 %
	 %
	 %
	 &= \nodePermutationMatrix{\nodeIndexOne}\,
	 \left(
	 \diag{ 0 \identityMatrix{}, \bogusPrecision{\nodeIndexOne}, \raisedInfty \identityMatrix{}}
	 +
	 \diag{ t_{\nodeIndexOne}
	 	\left(
	 	\begin{array}{ll}
	 	\observedVariance{\nodeIndexOne}  & \finiteObservedVariance{\nodeIndexOne} \\
	 	- & \finiteVariance{\nodeIndexOne} \\
	 	\end{array}
	 	\right), 0 \identityMatrix{}
	 }
	 \right)^{\specialInverse}
	 \nodePermutationMatrix{\nodeIndexOne}\transpose \\
	 %
	 %
	 %
	 %
	 &= \nodePermutationMatrix{\nodeIndexOne}\,
	 \left(
	 \diag{ \invertibleBlock
	 	,
	 	\raisedInfty \identityMatrix{}
	 }
	 \right)^{\specialInverse}
	 \nodePermutationMatrix{\nodeIndexOne}\transpose \\
	 &= \nodePermutationMatrix{\nodeIndexOne}\,
	 \diag{
	 	\invertibleBlock^{-1},
	 	0 \identityMatrix{}
	 }
	 \nodePermutationMatrix{\nodeIndexOne}\transpose,
	 \end{aligned}
	 \label{eq:fancyDeflation}
	 \end{equation}
	 where
	 \begin{equation}
	 \invertibleBlock =
	 \diag{ 0 \identityMatrix{}, \bogusPrecision{\nodeIndexOne}}
	 +
	 t_{\nodeIndexOne}
	 \left(
	 \begin{array}{ll}
	 \observedVariance{\nodeIndexOne}  & \finiteObservedVariance{\nodeIndexOne} \\
	 - & \finiteVariance{\nodeIndexOne} \\
	 \end{array}
	 \right)
	 = \left(
	 \begin{array}{cc}
	 t_{\nodeIndexOne} \observedVariance{\nodeIndexOne}  & t_{\nodeIndexOne} \finiteObservedVariance{\nodeIndexOne} \\
	 - & \bogusPrecision{\nodeIndexOne} + t_{\nodeIndexOne} \finiteVariance{\nodeIndexOne}
	 \end{array}
	 \right).
	 \end{equation}
	 The matrix $\invertibleBlock$ is the sum of a positive-definite matrix and positive-semidefinite matrix and is therefore invertible.

	 \section{Heritability Statistic}\label{app:heritability}
	 
	 We compute the expectation of the empirical variance $\expectedTotalVariance$ under the MBD model with residual variance as follows:
	 \begin{equation} \label{eq:expected_covariance_pre}
	 \begin{split}
	 \expectedTotalVariance &= \expectation{\frac{1}{\ntaxa} \left(\datamat - \matMean\right)\transpose \left(\datamat - \matMean\right)}  \\
	 &= \frac{1}{\ntaxa}\expectation{\datamat\transpose\datamat - \frac{2}{\ntaxa} \datamat\transpose\oneMatrix{\ntaxa}\datamat + \frac{1}{\ntaxa^2}\datamat\transpose\oneMatrix{\ntaxa}\oneMatrix{\ntaxa}\datamat}\\
	 &= \frac{1}{\ntaxa}\expectation{\datamat\transpose\datamat - \frac{2}{\ntaxa} \datamat\transpose\oneMatrix{\ntaxa}\datamat + \frac{1}{\ntaxa}\datamat\transpose\oneMatrix{\ntaxa}\datamat}\\
	 &= \frac{1}{\ntaxa}\expectation{\datamat\transpose\datamat - \frac{1}{\ntaxa}\datamat\transpose\oneMatrix{\ntaxa}\datamat} \\
	 &= \frac{1}{\ntaxa}\sum_{\taxonIdx = 1}^\ntaxa \expectation{\datamat_\taxonIdx\datamat_\taxonIdx\transpose} - \frac{1}{\ntaxa^2} \sum_{\taxonIdx = 1}^\ntaxa \sum_{\nodeIndexTwo = 1}^\ntaxa \expectation{\datamat_\taxonIdx \datamat_\nodeIndexTwo\transpose}.
	 \end{split}
	 \end{equation}
	 The multivariate normal distribution of $\vecOperator{\datamat}$ implies $\cov{\datamatsm_{\taxonIdx\traitIdxTwo}, \datamatsm_{\nodeIndexTwo\traitIdxThree}} = \facprecDiffsm_{\traitIdxTwo\traitIdxThree} \treeVarianceElement_{\taxonIdx\nodeIndexTwo} + \samplingVarsm_{\traitIdxTwo\traitIdxThree}\indicator{\taxonIdx}{\nodeIndexTwo}$ where $\indicator{\taxonIdx}{\nodeIndexTwo}$ is an indicator function.
	 Using this information in SI Equation \ref{eq:expected_covariance_pre},
	 \begin{equation}
	 \begin{split}
	 \expectedTotalVariance &= \frac{1}{\ntaxa}\sum_{\taxonIdx = 1}^\ntaxa \left( \treeVarianceElement_{\taxonIdx\taxonIdx} \facprecDiff + \samplingVar + \expectation{\datamat_\taxonIdx}\expectation{\datamat_\taxonIdx}\transpose\right) \\
	 &\hspace{1cm}- \frac{1}{\ntaxa^2} \sum_{\taxonIdx = 1}^\ntaxa \sum_{\nodeIndexTwo = 1}^\ntaxa \left( \treeVarianceElement_{\taxonIdx\nodeIndexTwo} \facprecDiff + \samplingVar\indicator{\taxonIdx}{\nodeIndexTwo} + \expectation{\datamat_\taxonIdx}\expectation{\datamat_\nodeIndexTwo}\transpose\right) \\
	 &= \frac{1}{\ntaxa}\trOperator{\treeVariance}\facprecDiff + \samplingVar - \left(\frac{1}{\ntaxa^2} \oneVector{\ntaxa}\transpose \treeVariance \oneVector{\ntaxa}\right)\facprecDiff - \frac{1}{\ntaxa}\samplingVar \\
	 &\hspace{1cm} + \frac{1}{\ntaxa}\sum_{\taxonIdx = 1}^\ntaxa \expectation{\datamat_\taxonIdx}\expectation{\datamat_\taxonIdx}\transpose - \frac{1}{\ntaxa^2} \sum_{\taxonIdx = 1}^\ntaxa \sum_{\nodeIndexTwo = 1}^\ntaxa \expectation{\datamat_\taxonIdx}\expectation{\datamat_\nodeIndexTwo}\transpose.
	 \end{split}
	 \end{equation}
	 Note that $\expectation{\datamat_\taxonIdx} = \datamat_\nroot$ for $\taxonIdx = 1 \hdots \ntaxa$, which implies
	 \begin{equation}
	 \frac{1}{\ntaxa}\sum_{\taxonIdx = 1}^\ntaxa \expectation{\datamat_\taxonIdx}\expectation{\datamat_\taxonIdx}\transpose - \frac{1}{\ntaxa^2} \sum_{\taxonIdx = 1}^\ntaxa \sum_{\nodeIndexTwo = 1}^\ntaxa \expectation{\datamat_\taxonIdx}\expectation{\datamat_\nodeIndexTwo}\transpose = 0.
	 \end{equation}
	 As such, our expression for the expected empirical variance reduces to the following:
	 \begin{equation}
	 \expectedTotalVariance = \frac{\ntaxa - 1}{\ntaxa} \samplingVar  + \left(\frac{1}{\ntaxa} \trOperator{\treeVariance} - \frac{1}{{\ntaxa}^2}\oneVector{\ntaxa}\transpose \treeVariance  \oneVector{\ntaxa}\right)\facprecDiff.
	 \end{equation}
	 
	 \section{Algorithm for Efficiently Computing $\heritabilityMatrix$}\label{app:treeSum}
	 Note that naive computation of $\diffConstant = \frac{1}{\ntaxa} \trOperator{\treeVariance} + \frac{1}{{\ntaxa}^2}\oneVector{\ntaxa}\transpose \treeVariance  \oneVector{\ntaxa}$ in Equation \ref{eq:CoheritabilityStat} would require constructing the $\ntaxa \times \ntaxa$ matrix $\treeVariance$ and summing over all its elements, which has computation complexity of at least $\bigo{\ntaxa^2}$.
	 For cases where $\treeparam$ is random and changes throughout the MCMC simulation, this quantity must be re-computed each time we compute the statistic.
	 To avoid this issue, we implement an algorithm that avoids constructing $\treeVariance$ in its entirety and simply calculates both $\trOperator{\treeVariance }$ and $\oneVector{\ntaxa}\transpose \treeVariance \oneVector{\ntaxa}$ in $\bigo{\ntaxa}$ time.
	 The algorithm performs a post-order traversal of the tree where at each internal node $\node_\parentIdx$ we compute $\nBelow{\parentIdx}$ (the number of tips below $\node_\parentIdx$), $\sumBelow{\parentIdx}$ (the sum of all elements in $\treeVarBelow{\parentIdx}$), and $\diagBelow{\parentIdx}$ (the sum of the diagonal elements in $\treeVarBelow{\parentIdx}$).
	 We define $\treeVarBelow{\parentIdx}$ as the tree variance-covariance matrix constructed from the sub-tree $\treeBelow{\parentIdx}$ that is simply the tree that contains only the nodes below $\node_{\parentIdx}$ with node $\node_\parentIdx$ as its root.
	 For internal nodes $\node_\parentIdx$ with child nodes $\node_\childIdxOne$ and $\node_\childIdxTwo$, we accumulate
	 \begin{equation}
	 \begin{split}
	 \nBelow{\parentIdx} &= \nBelow{\childIdxOne} + \nBelow{\childIdxTwo} + 1,\\
	 \sumBelow{\parentIdx} &= \sumBelow{\childIdxOne} + \sumBelow{\childIdxTwo} + \branchLength{\childIdxOne}\nBelow{\childIdxOne}^2 + \branchLength{\childIdxTwo}\nBelow{\childIdxTwo}^2,\text{ and}\\
	 \diagBelow{\parentIdx} &= \diagBelow{\childIdxOne} + \diagBelow{\childIdxTwo} + \branchLength{\childIdxOne}\nBelow{\childIdxOne} + \branchLength{\childIdxTwo}\nBelow{\childIdxTwo}.
	 \end{split}
	 \end{equation}
	 At the tips, we initialize with $\sumBelow{\taxonIdx} = \diagBelow{\taxonIdx} = 0$ and $\nBelow{\taxonIdx} = 1$.
	 At the root, $\sumBelow{2\ntaxa - 1} = \oneVector{\ntaxa}\transpose \treeVariance  \oneVector{\ntaxa}$ and $\diagBelow{2\ntaxa - 1} = \trOperator{\treeVariance }$. This algorithm visits each node in $\treeparam$ exactly once and has run time $\bigo{\ntaxa}$.

\end{document}